# Thanks to 2D and maybe even beyond:
# 115 GeV and 140 GeV almost Standard Model Higgs without problems

**Marko B Popovic \***

**(Cambridge, MA, July 24, 2010)**

**Abstract**

I address four key points: (1) dynamical, likely *local* Higgs mass generation (potentially in 2D) as resolution to the 4D hierarchy and vacuum energy problems, (2) possibility that top condensation may be explained by an interplay among the gluon and scalar sectors, (3) the Higgs Mass Zero Crossing (HMZC) scale, most likely equal to the electro-weak symmetry breaking scale, $\Lambda_{EWSB}$, in accord with standard cosmology or classic inflation, and (4) two preferred Higgs regions centered at $116.5\ GeV$ and $140.5\ GeV$ with related high energy models.

I show that the Standard Model (SM) in 2D could simultaneously satisfy (a) complete radiative generation of the Higgs mass via top loop and (b) cancelation of the remaining leading order corrections to the scalar propagator. The Higgs mass, parameterized with k=1 (2), in the leading order is $113.0 \pm 1.0\ GeV$ ($143.4 \pm 1.3\ GeV$).

I show that the SM top condensation is consistent with the gluon and Higgs mediated top-anti top interactions at tree level. I predict the QCD fine structure constant up to precision better than 2% with the mean value surprisingly only 0.25% away from the world average value at $\sqrt{s} = M_Z$.

The SM driven theory at energies larger than the HMZC scale necessary includes effective tachyonic Higgs (Popovic 2001). Here, I map the SM physical Higgs mass to the low energy HMZC scale ($0.8 - 1.8\ TeV$). I show that the *very* "long lived" SM, valid up to the Planck scale, necessitates (a) Higgs heavier than $137.0 \pm 1.8\ GeV$ due to the vacuum stability, (b) Higgs lighter than $171 \pm 2\ GeV$ due to the perturbativity and (c) Higgs lighter than $146.5 \pm 2\ GeV$ such that there is a single HMZC scale at energies smaller than the Planck mass.

I present candidate $m_H = 138.1 \pm 1.8\ GeV$ for the SM valid up to an energy scale, nearly equal Planck mass, obtained from a conjecture which minimizes the parameters of the Higgs potential. I show that the Minimal Supersymmetric SM is less unnatural than the SM for $m_H \leq 120.9 \pm 0.9\ GeV$. I introduce a class of models potentially *exactly* removing tachyons. I analyze class of the Composite Particles Models (CPM) (Popovic 2002) where top quark is composite, composed of 3 fundamental fermions, and Higgs scalar is composite, composed of 2 fundamental fermions, with $m_H \cong \frac{2}{3} m_t = 115.4 \pm 0.9\ GeV$.

(52 Pages, 11 Figures, 1 Table, 155 References)

*Please address all correspondence to:
Marko B Popovic, PhD,
115 Harvard St #3, Cambridge, MA 02139, USA
Email: mbpopovic@gmail.com ;
URL:   http://web.media.mit.edu/~marko/ ,
       http://walltrust.com/MarkoBPopovic/index.htm

# 1. Introduction

The Standard Model (SM) [1-21] of particle physics has been verified during the last four decades with an unprecedented accuracy. The Higgs scalar particle is the last SM ingredient that has still not been experimentally confirmed. Compared to all other particles Higgs is expected to be a very unique one.

Higgs is anticipated to acquire a non-zero *vacuum expectation value* (VEV) in the early Universe when the average collision center of mass energies reach certain energy scale $\Lambda_{EWSB}$. This VEV then "breaks" the original electroweak symmetry mediated by four types of lights, one associated with the "hypercharge" gauge symmetry and another three associated with the "weak" gauge symmetries. As the Universe cools down photon remains massless while $Z, W^{\pm}$ gauge bosons acquire masses.

Moreover, by coupling with itself and other particles Higgs is expected to generate its mass as well as masses for all other known massive particles. The Higgs particle is, therefore, expected to bear most of the responsibility for the mass generation in the known Universe.

Higgs is indirectly or directly anticipated to be the root of majority of physical phenomena. It is no surprise that some popular media call it the "God particle". In a sense, Higgs is a modern times version of the *ether* concept that was, hence, just partially removed by the Einstein's Special Relativity [22].

The Large Electron Positron (LEP) particle accelerator near Geneva observed a number of suspicious events [23-24] in the vicinity of $115 \, GeV/c^2$, at the center of mass energies a bit above $\sqrt{s} \cong 206 \, GeV/c^2$, just before the accelerator was shut down in 2000. Now, after 10 years, the LEP's successor, the Large Hadron Collider (LHC), [25] the biggest ever endeavor in the particle physics research, is again collecting high energy data (on the order of $7 \, TeV$). As I discuss later on, it is quite certain that new physics phenomena, beyond current physics dogma [1-21], will be observed relatively soon.

First, I give an overview of the traditional set of problems associated with the SM Higgs scalar particle: hierarchy and vacuum energy. I then address the renormalized SM at high energies; if the SM is valid and complete description at given energies, under a few reasonable assumptions, the SM high energy behavior may pose important limits to the SM physical Higgs mass. In this paper, I carefully map the physical Higgs mass with the low energy SM Higgs Mass Zero Crossing (HMZC) scale $\sim \Lambda_{EWSB}$ [26] at which the effective Higgs mass is zero; the effective SM Higgs particle goes from regular massive particle $m_H^2 \geq 0$ at small energies to tachyon degree of freedom $m_H^2 < 0$ at large energies. Also, because there are two HMZC branches per Higgs mass, only the branch with a correct crossing should be considered affiliated with the scale of the electroweak symmetry breaking $\Lambda_{EWSB}$. I explain why I expect $\Lambda_{HMZC} \sim \Lambda_{EWSB}$ to be in accordance with the standard cosmology or classic inflation and why the tachyonic Higgs in the broken phase matches the non-tachyonic, scalar in the *very* early Universe.

In Section 2, I obtain the Higgs mass range for the Minimal Supersymmetric SM [27-29], see also [30-31], for which the MSSM is less unnatural theory than the SM at low energies.

In Section 3, I explain why two particular Higgs mass regions, centered at $113 \, GeV$ and $143 \, GeV$, may be favored on the theoretical grounds based on an analysis of the leading quantum corrections in 2D.

Both hierarchy and vacuum energy problems may be eradicated If EW symmetry breaking takes place in 2D which "embeds" the physical propagation in 4D. The leading divergences in 2D are only logarithmically divergent and therefore the large hierarchy may be easily attained. Moreover, the Higgs scalar field doesn't need to be non-zero constant in the entire 4D space hence leading to reasonable 4D space-time curvature. As addressed later on, there may be different forms of "embedding" 2D into 4D. Hence, I hypothesize that standard 4D EWSB might be only an useful description that corresponds to dynamical, effective and potentially *local* 2D description of EWSB.

Finally, independent of the space-time dimensionality, I suggest that the EW symmetry breaking may be both dynamical and *local* in sense that effective Higgs scalar field might be zero in almost entire space except in the close vicinity of physically propagating massive particles.

In Section 4, I give a brief overview of the previous work related to dynamical top quark condensation.

In Section 5, I show that top anti-top condensate formation may be consistent with interplay between the QCD gluon and Higgs mediated top interactions. This is good news, as this symmetry breaking defining, or maybe just contributing principle, may span vast energy scales in a natural fashion. This analysis should not be confused with analysis that leads to the QCD driven top quark infrared fixed point, predicting top quark mass near $\sim 230\ GeV$, e.g. see [32], and, as emphasized by Nambu, physical, observable, low energy Higgs scalar bound state with a mass of $m_H \cong 2m_t$ to leading order in $1/N_C$.

In Section 6, I discuss "smooth" transitions across the energy scale $\Lambda_{HMZC} \sim \Lambda_{EWSB}$ where effective theory does not abruptly change the parameters and degrees of freedom. Hereafter I dub this transition Type II. I explain why the hierarchy problem is actually a rather benign problem within the minimal SM with composite Higgs in 4D at energies larger than the electroweak breaking scale; i.e. theory may span across vast energy scales. I illustrate this high energy SM behavior with particularly interesting Higgs mass candidate, $\sim 138\ GeV$, where the SM is assumed to be valid up to a composite energy scale, nearly equal Planck mass, obtained from a conjecture that minimize the parameters of the Higgs potential [33]. Moreover, I address the Planck scale adaptation of the Coleman-Weinberg (CW) conjecture [34].

In Section 7, I discuss "discontinuous" transitions across the energy scale $\Lambda_{HMZC} \sim \Lambda_{EWSB}$ where effective theory abruptly changes the parameters and degrees of freedom. Hereafter I dub this transition Type I. I also sketch several simplified models; these models are mainly related to the top quark sector and they deal with external particle degrees of freedom within 2D and 4D space-times as well as with degrees of freedom (color, flavor etc) within the internal space. I introduce a class of models that potentially may *exactly* remove the tachyon solution at high energies. I analyze class of models, see [33], where top quark is composite, composed of 3 fundamental fermions, and Higgs scalar is composite, composed of 2 fundamental fermions, with $m_H \cong \frac{2}{3} m_t = 115.4 \pm 0.9\ GeV$.

While I expect Type I transition to be most likely identified with the first order EW phase transition, Type II transition, in principle, may correspond to either the first or the second order EW phase transition.

In the conclusion, I summarize and discuss findings as well as present the best Higgs mass candidates in the vicinity of $116.5\ GeV$ and $140.5\ GeV$; these are in good agreement with the predictions in [33].

I also "prove" that the LHC is already making its mark in history by stepping distinctively outside of the region covered by well-established phenomenology. Even if the SM is valid description below and above the HMZC scale, never in the past have particle physicists dealt with an effective theory that includes tachyons, i.e. particles with negative effective mass squared. These are exactly the characteristic of the SM driven theory above the HMZC scale unless there is some other yet unknown physics. While tachyon theories are commonly addressed in the context of string theory and cosmology there is an *alarming lack* of literature and ongoing research effort among the rest of physics community.

As I discuss later on, this "proof" should not be confused with the perturbative unitarity considerations [35,36] which motivate the existence of Higgs particle at low energies but not necessarily the new physics; hence, existence of the HMZC scale is better stimulus for the new physics beyond the SM Higgs.

In this manuscript I tried to present material with enough clarity and detail so that graduate physics students in their first years and maybe even a few advanced senior physics major students could easily follow, understand, and reproduce all major points. I apologize to reader if I failed to succeed in my goal.

## 2. Current state of affairs
### 2.1. Problems with current model

Traditionally, there are two main problems with the SM Higgs model: (1) *Hierarchy* (or fine tuning problem/naturalness) and (2) *Vacuum energy problem*.

**Hierarchy** is usually associated with the idea that there are likely two important energy scales separated by many orders of magnitude. One is the electroweak symmetry breaking scale $\Lambda_{EWSB}$, expected to be on the order of magnitude of the EW VEV, $v_{EW} = 2.462 \cdot 10^2 \, GeV$; as I discuss later on, the $\Lambda_{EWSB}$ is more closely related to the low energy HMZC scale $\sim 10^3 \, GeV$ than to VEV, see Section 2.3. The other important energy scale is the Planck mass energy scale, $M_{Pl} \sim 10^{19} \, GeV$, at which quantum physics is traditionally expected to become strongly entangled with gravity, i.e. dynamics of the 4D curved space-time described by the General Relativity [37,38]. The hierarchy problem is how to connect these two largely separated scales within single meaningful theoretical framework, expressed in the spirit of the effective theory and Wilson's approach to renormalization theory [39].

Traditionally, one of the main obstacles is the presence of divergences (without the high-energy cut-off that would be infinities) that grows quadratically with energy scale. The scale-renormalized Higgs mass in 4D grows quadratically (see Appendix) and if Higgs mass in the vicinity of each of the two important scales is expected to be on the order of that energy scale the parameters of the theory might need to be *fine-tuned*. A slight change of the parameters at one scale causes large changes at the other scale, see Fig 4. Clearly, if there is fine-tuning [40] without explanation then such a model is considered *unnatural*.

**Vacuum energy problem** is among other caused by the non-zero VEV of the Higgs scalar field traditionally expected to span the entire 4D space-time. This, however, implies a huge energy density everywhere and, hence, an enormously large space-time curvature. Therefore, the Higgs mechanism is by many orders of magnitude inconsistent with our everyday physical reality [41-42].

Similarly, if the Universe is described by an effective local quantum field theory down to the Planck scale, then one would expect a cosmological constant of the order of $M_{Pl}^4$. The measured cosmological constant is smaller than this by a factor of $10^{-120}$. This discrepancy is termed "the worst theoretical prediction in the history of physics!" [43].

### 2.2. The SM Higgs Mass: Direct Searches and Indirect limits from the electroweak precision data

Back in 2000, based on LEP2 data, ALEPH reported an excess of about three standard deviations, suggesting the production of a SM Higgs boson with mass $\sim 115\ GeV$ [44]. The combined analysis by ALEPH, DELPHI, L3, and OPAL could not either confirm or exclude the $\sim 115\ GeV$ Higgs; instead, that analysis placed a current 95% C.L. lower bound of $114.4\ GeV$ for the mass of the SM Higgs boson based on direct searches [45]; see Particle Data Group review [46] and references therein.

A global fit to the precision electroweak data, accumulated in the last decade at LEP, SLC, Tevatron and elsewhere, gives $m_H = 76 + 33 - 24\ GeV$, or $m_H < 144\ GeV$ at 95% C.L. [47]. However if the direct LEP search limit of $m_H > 114.4\ GeV$ is taken into account, an upper limit of $m_H < 182\ GeV$ at 95% C.L. is obtained for the SM Higgs mass [46]. Finally, stringent limits, $114\ GeV < m_H < 145\ GeV$ at 95% C.L. are obtained from the revisited global electroweak fit [48].

Therefore, by anticipating results presented in Sections 2.3 and 2.4., global fit and direct searches suggest that the SM Higgs is within the HMZC regime, see also Fig 4.

The recent Tevatron analysis [49-51] excluded the Higgs within the $162\ GeV < m_H < 166\ GeV$ range at 95% C.L.; whereas the entire $160\ GeV < m_H < 170\ GeV$ range appears to be unlikely. Finally, the combined CDF and DØ analysis [49] observed $1\sigma$ excess beyond the SM expected background for the $132\ GeV < m_H < 143\ GeV$ Higgs range. This is the result of the log-likelihood ratio analysis of the $4.8\ fb^{-1}$ and $5.4\ fb^{-1}$ integrated luminosities data collected with the CDF and DØ detectors, respectively.

### 2.3. The SM at high energies as function of the Higgs mass

The effective parameters of the SM theory may be explored at high energies according to the renormalization group flow, i.e. according to the effective theory and Wilson's approach to renormalization theory [39]. The parameters of practical importance here are gauge couplings, top Yukawa coupling, and finally, parameters defining Higgs potential energy density

$$V = -\frac{m_H^2}{4}|\Phi|^2 + \frac{\lambda}{8}|\Phi|^4 \qquad (1)$$

where $m_H^2$ is the Higgs mass squared and $\lambda$ is the Higgs scalar quartic coupling. This analysis is meaningful only within the energy range where the SM may be considered valid and complete effective theory.

Here, I overview previous results [26, 33] on the vacuum stability [52-59], the perturbativity [60-61] and the Higgs Mass Zero Crossing (HMZC) scale [26, 33] associated with the electroweak phase transition.

When the SM is run up in energies the effective Higgs goes from regular massive particle $m_H^2 \geq 0$ at small energies to tachyon degree of freedom $m_H^2 < 0$ at large energies. Transitional scale, dubbed the HMZC scale, represents a more "pedantic" estimate of the EWSB scale [26] than provided by the scalar VEV; HMZC scale is discussed in more details in the Section 2.4.

Generally, there are two, one low and one high energy, HMZC branches per physical Higgs mass and only the low energy branch with a correct crossing $m_H^2 \geq 0, \Lambda < \Lambda_{HMZC} \to m_H^2 < 0, \Lambda > \Lambda_{HMZC}$, should be considered affiliated with the scale of the electroweak symmetry breaking $\Lambda_{EWSB}$.

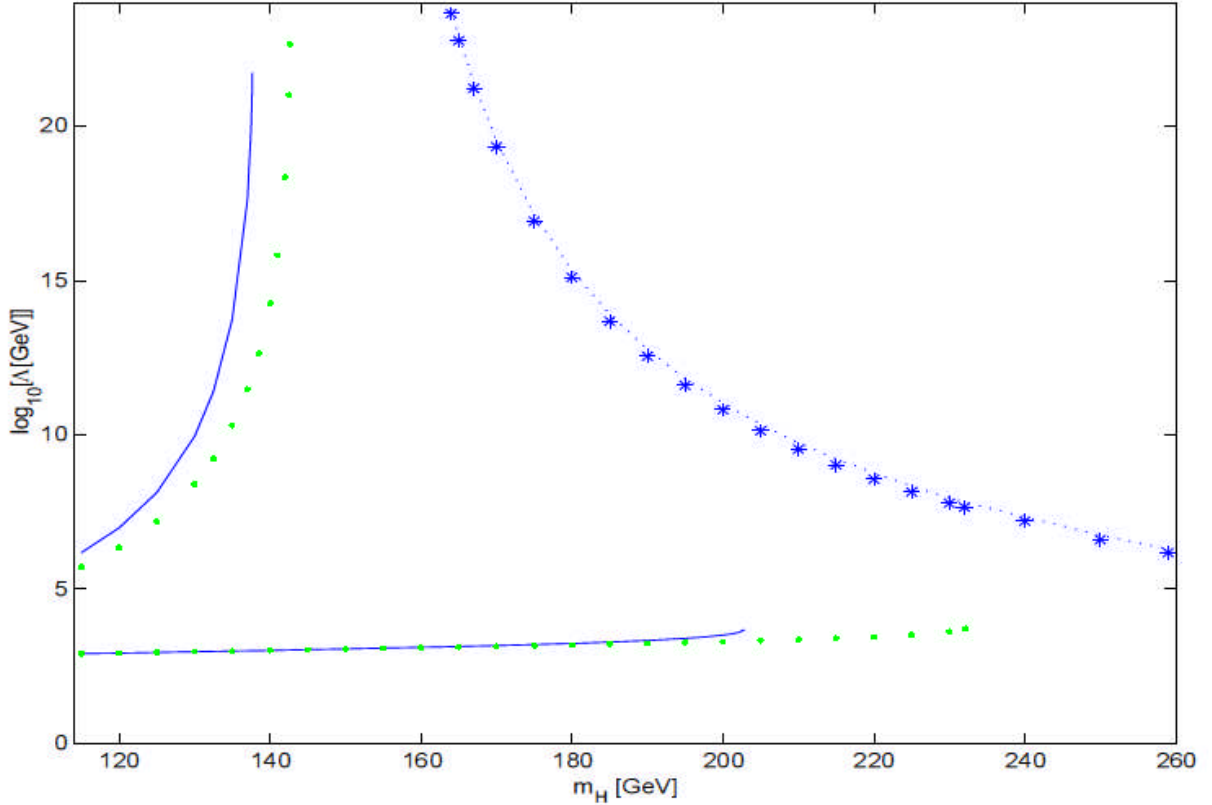

**Figure 1** **The SM high energy scales as function of the SM Higgs mass [26]. The vacuum stability limit (upper left corner) as obtained in the hard-cutoff method (dotted) and the $\overline{MS}$ scheme (solid). The perturbativity limit (upper right corner) as obtained in the hard-cutoff method (dotted) and the $\overline{MS}$ scheme (solid). The HMZC scale [26] (bottom line), $\sim\Lambda_{EWSB}$, as obtained in the hard-cutoff method (dotted) and the $\overline{MS}$ scheme (solid).**

The results presented in Section 2.3.1 are solutions to an ordinary first order differential equation with boundary condition set by the physical Higgs mass. The running of all quantities of interest is smooth; hence, the obtained results, at least in the numerical sense, are expected to be valid up to very high energies. The results are mainly sensitive to the top quark mass, see Section 2.3.1. Two independent techniques were utilized in parallel: (1) the $\overline{MS}$ scheme [62], applied to the effective potential [34] analysis [63-65], and (2) the Euclidean hard cut-off scheme, applied to the generalized original Veltman's approach [66], subsequently confirmed by Osland and Wu [67] and completed with the logarithmic divergences by Ma [68]. Details of this analysis are provided in Appendix; more details are provided in the original publication [26] and references therein.

### 2.3.1. **Numerical results**

For the Higgs mass smaller than $137.0 \pm 1.8\ GeV$ [26] there is an energy scale smaller than the Planck scale at which unacceptable deeper minima of the SM effective potential occur. This is usually referred to as the stability criteria [52-59]. For the Higgs mass larger than $171 \pm 2\ GeV$ [26] there is an energy scale smaller than the Planck scale at which the Higgs scalar quartic coupling $\lambda$ reaches the Landau pole (essentially blows up) and the Higgs scalar sector becomes strongly coupled. This is usually referred to as the perturbativity criteria [60-61]. A more conservative estimate may include an additional $o(3\ GeV)$ uncertainty [26] for the Higgs mass, in response to the requirement that effective potential must be renormalization scale independent; see Appendix for further details. Finally, for the Higgs mass smaller than $203^{+14}_{-3}\ GeV$ [26] the HMZC scale exists at which the renormalized effective Higgs mass is zero.

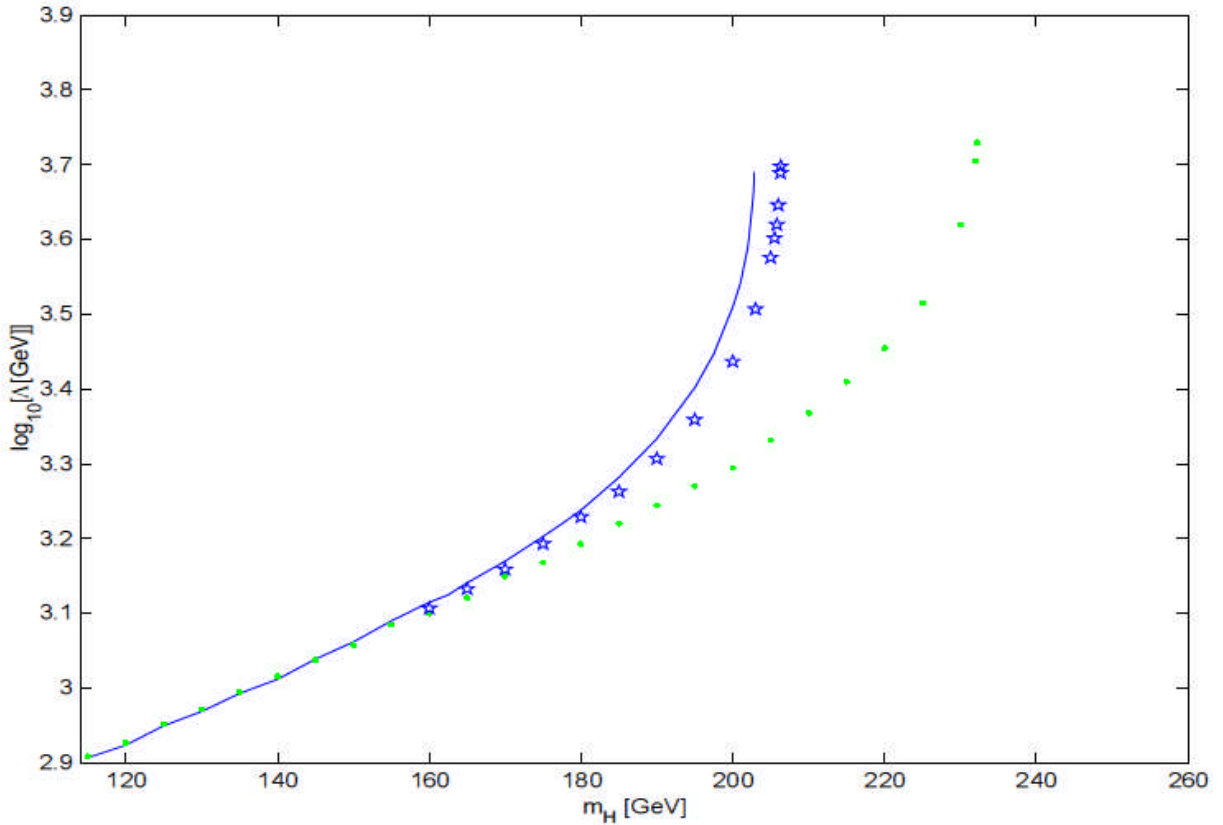

**Figure 2** The Higgs Mass Zero Crossing (HMZC) scale as function of the SM physical Higgs mass in the hard-cutoff method (dotted), the $\overline{MS}$ scheme (solid) and with matching as in [61] (pentagrams) [26].

The stability, perturbativity and low energy HMZC curves are shown in Fig 1-2 and Higgs mass "running" for several Higgs masses is shown in Fig 3 [26, 33]. This was obtained with $m_t = 175\ GeV$ and $\alpha_S(M_Z) = 0.1182$; see [26]. Due to variability [26] in the Higgs mass $\delta m_H[GeV] \cong 1.4\ \delta m_t[GeV] - 360 \delta \alpha_S$, the current world average top quark mass from combined CDF and DØ analysis, $m_t = 173.1 \pm 1.3\ GeV$ [69], introduces a small shift and implies existence of the HMZC scale for the Higgs mass below $200.3^{+14}_{-3}\ GeV$.

The low energy HMZC branch is in the range 800 $GeV$ and 5 $TeV$ and high energy HMZC branch is larger than the Planck mass for the physical Higgs mass smaller than $146.0 \pm 2.0\ GeV$ (33). The variability in the Higgs mass, here $\delta m_H [GeV] \cong -0.24\ \delta m_t [GeV]$, introduces a small shift, and the upper limit on the physical Higgs mass is now $146.5 \pm 2.0\ GeV$. Again, a more conservative estimate may include additional $o(3\ GeV)$ uncertainty [26] for the Higgs mass, in response to the requirement that effective potential must be renormalization scale independent, see Appendix.

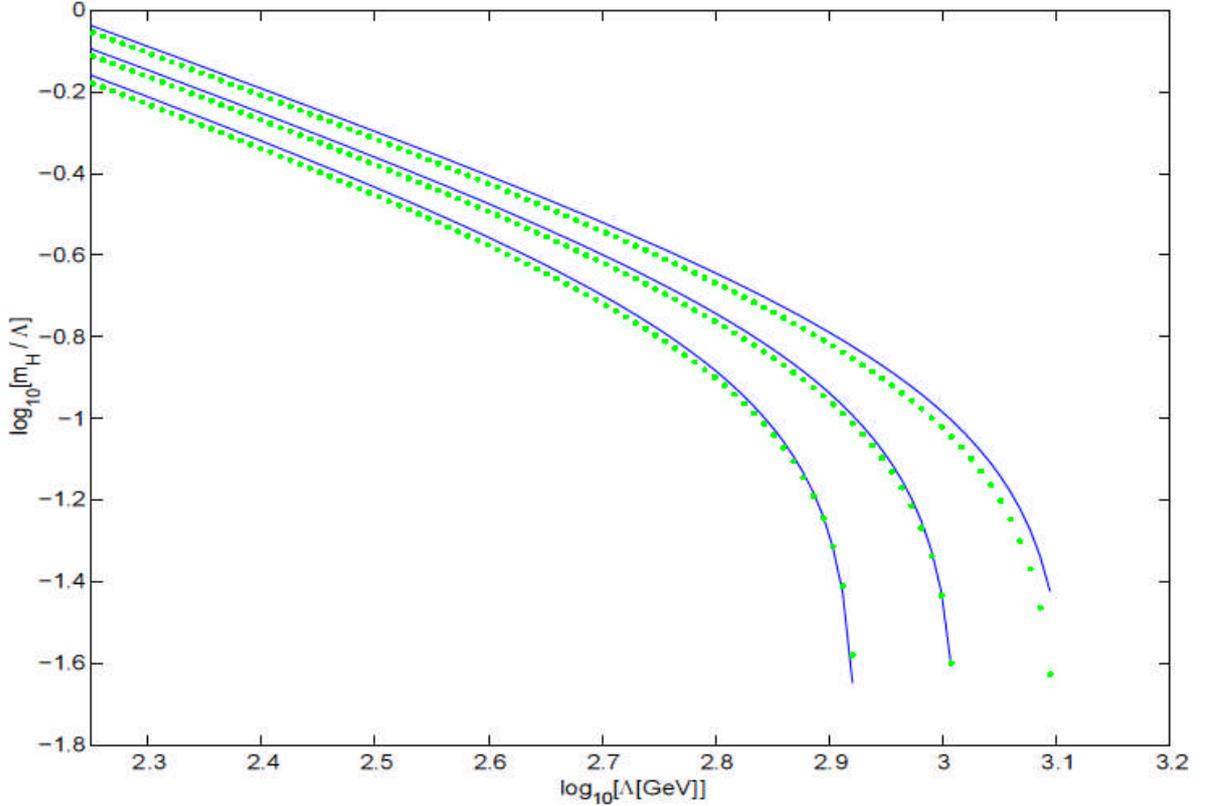

**Figure 3** **The logarithm of the effective Higgs mass rescaled with energy as function of the energy scale for several SM physical Higgs masses (from left to right 120, 140 and 160 GeV) in the hard-cutoff method (doted) and in the $\overline{MS}$ scheme (solid).[26]**

Therefore, if the SM is a valid description of Nature all the way up to the Planck scale, where effective potential corresponds to an unbroken electroweak symmetry, then stability curve and condition requiring a single HMZC below the Planck scale limit the Higgs masses to a very tight window of roughly $142 \pm 6\ GeV$ with an electroweak phase transition scale roughly in the range $1 - 1.15\ TeV$.

However, Type I transition, taking place at the HMZC scale, may introduce an abrupt change of the parameters and the degrees of freedom. Therefore, there may be two very different descriptions below and above the HMZC scale that could make the above considerations inappropriate for energies larger than the low energy HMZC scale $\sim \Lambda_{EWSB}$.

2.4. **Significance of the SM HMZC scale?**

**Q:** What are the physical significances of the HMZC scale?

**A:** This is an energy scale that separate effective standard Higgs particle from tachyonic Higgs particle; hence, any "new physics" that "asymptotically" tends to or approximates the SM needs to explain the positive or negative Higgs mass squared for energies smaller or larger than the HMZC scale respectively.

Therefore, the HMZC scale may serve as useful reference point for any type of new physics model building that aspire to smoothly approximate the SM at low energies and low temperatures. As I discuss later on, the HMZC scale may also serve as useful reference even for the more dramatic transitions.

The probable meaning of the HMZC scale in the context of the *very* early Universe is addressed below.

**Q**: Is the low energy $\Lambda_{HMZC}$ scale affiliated with the electroweak symmetry breaking, $\Lambda_{EWSB}$, scale?

**A**: Most likely, unless the effective theory was dominantly tachyonic at some point in the *very* early Universe. Traditionally the EWSB scale corresponds to the CM energies in the early Universe when first (Type I or II) or second (Type II) order EW phase transition took a place. The order parameter is the temperature dependent vacuum expectation value. If there is no time period when the early Universe is dominated by the effective tachyon physics then, necessary, $\Lambda_{HMZC}$ is to be identified with $\Lambda_{EWSB}$.

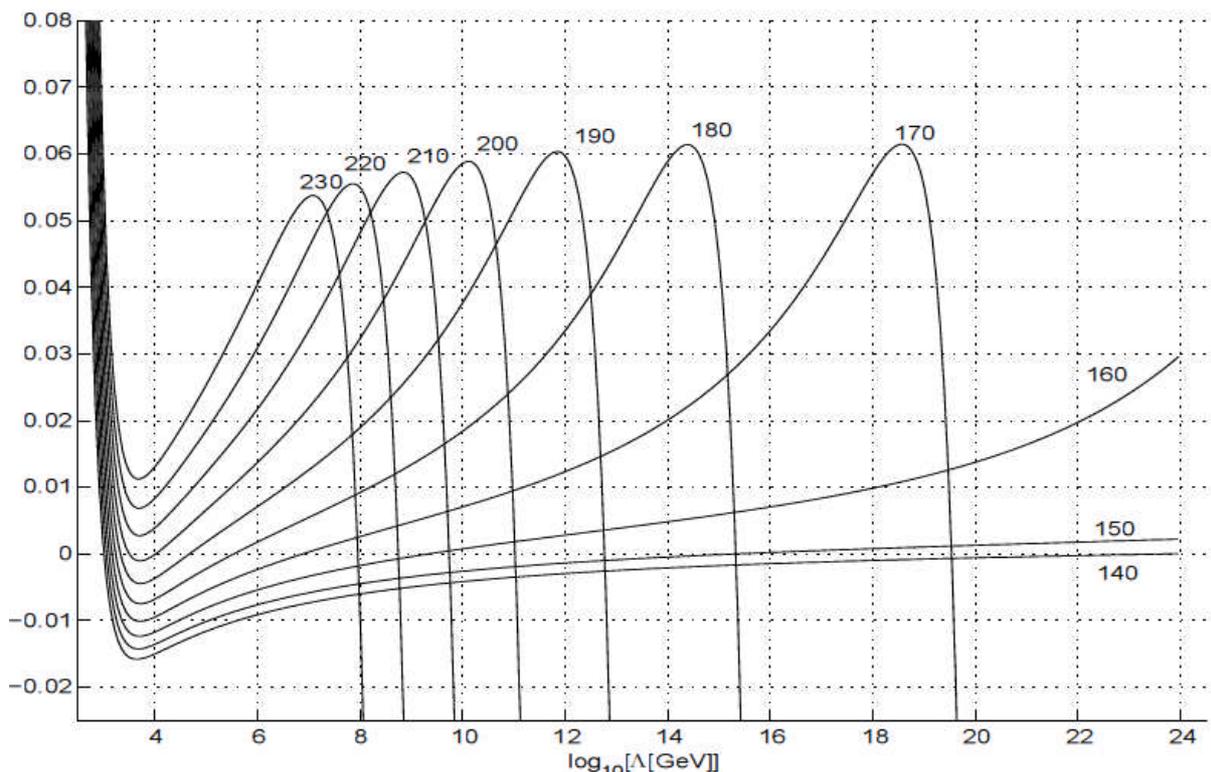

**Figure 4** The SM dimensionless parameter $\mu$, i.e. effective Higgs mass squared divided by the energy scale squared, as function of the energy scale for ten physical Higgs masses (from left to right 140, 150, 160, 170, 180, 190, 200, 210, 220 and 230 GeV) in dimensional $\overline{MS}$ regularization.[33]

The electroweak symmetry breaking is thought of here as a phase transition at non-zero temperature. By preparing system the average CM collision energies can be brought to the HMZC energy $\sim\Lambda_{\text{EWSB}}$, which then creates a condition for the phase transition, which probably happened in the early Universe [70]. Hence, by going back in time, the ground state of today's Universe (low temperature) with non-zero VEV (e.g. corresponding to the ordinary SM within the *Type II transitions)*, transitions to ground state with zero VEV (high temperature). In the hot, zero-VEV Universe there are no tachyons at short distances, i.e. energies larger than the HMZC energy, but rather just non-tachyonic scalar particle. As discussed in Section 2.6 the reason for that is different ground state!

The tachyon, however, exists in the zero-VEV Universe, for energies smaller than the HMZC scale (at large distances); but its effect on local dynamics is rather negligible because the average CM collision energies are much larger than the HMZC energy $\sim\Lambda_{\text{EWSB}}$.

Bottom line, if $\Lambda_{\text{HMZC}}\sim\Lambda_{\text{EWSB}}$, the dominant effective dynamics in the Universe is *always* described with non-tachyonic scalar particle and it is the ground state that changes when the average CM collision energy is equal to the HMZC energy or what is traditionally referred to as the EWSB scale! Alternatively, if $\Lambda_{\text{HMZC}} \neq \Lambda_{\text{EWSB}}$ then the standard hot Big Bang cosmology likely requires revision/augmentation. The tachyonic Higgs phase could probably explain inflation and more but this will be addressed elsewhere.

Hence in the former case, by going forward in time, as the temperature cools down just after the Big Bang, the ordinary scalar particle, in the zero-VEV Universe, transitions, for energies larger than the HMZC scale (short distances), to the tachyonic Higgs in the non-zero VEV Universe. For energies smaller than the HMZC scale (large distances), as the temperature cools down, the tachyonic scalar, in the zero-VEV Universe, transitions to the ordinary Higgs, in the non-zero VEV Universe.

Clearly, one necessitates details of the UV completion to describe details of the EW phase transition within the Type I transition which probably corresponds to the first order phase transition.

**Q:** Is the size of theoretically expected Higgs mass consistent with existence of the HMZC scale?

**A:** Yes. The loss of perturbative unitarity [35] without the Higgs sector occurs for the SM $W_L^+W_L^- \to W_L^+W_L^-$ s-wave scattering, see [71] for review, for $\sqrt{s} \approx \Lambda = 4\pi v_{EW} \cong 3\ TeV$. A slightly stronger bound, $\sqrt{s} \leq 2\sqrt{2}\pi v_{EW} \cong 2.2\ TeV$, is obtained by including the effect of the channel $W^+W^- \to ZZ$, see [35, 71]. Finally the naturalness considerations, see for example [36], suggest 1 TeV as the upper scale for either the SM Higgs or new physics that could play the role of Higgs sector at low energies. These bounds are in a very good agreement with the HMZC scale as obtained here.

**Q:** Is the size of the experimentally expected Higgs mass consistent with existence of the HMZC scale?

**A:** The global fit and direct searches, see Section, leads to an upper limit of $m_H < 182\ GeV$ at 95% C.L. [46], whereas revisited global electroweak fit [48] leads to an upper limit of $m_H < 145\ GeV$ at 95% C.L. , i.e. much narrower than the above theoretical limits. Hence, the expected SM Higgs is exactly within the range expected by the HMZC regime, see Fig 4.

**Q:** Does the HMZC scale suggest new physics?

**A:** Yes. The tachyonic Higgs certainly represents a "new physics". Similarly, if one believes that tachyonic Higgs should not exist then the HMZC scale must be considered as an upper bound on the scale of new physics without the tachyonic Higgs. Hence, the HMZC scale is better motivation for new physics than the perturbative unitarity consideration [35] which only motivates existence of standard Higgs sector at low energies but it does not necessarily motivate *new physics beyond the regular SM* Higgs particle.

### 2.5. The SM HMZC scale and the preferred Higgs mass for the Minimal Supersymmetry

The HMZC scale can distinguish a more meaningful from a less meaningful Minimal Supersymmetric Standard Model (MSSM) [27-29], see also [30-31], at low energies. By using the approximate relation [72-73] for the radiatively corrected Higgs mass in the MSSM, $m_H^2 \leq M_Z^2 + \frac{3G_F}{\sqrt{2}\pi^2} m_t^4 ln\left(\frac{m_T^2}{m_t^2}\right)$, one finds that the MSSM decoupling scale, defined here by the stop mass, $m_T$, is smaller than the HMZC scale for Higgs lighter than $m_H \leq 127.0\ GeV$ [26] for top quark mass $m_t = 175\ GeV$; this translates to $125.9\ GeV$ for the current world average $m_t = 173.1\ GeV$ [69] and $124.9\ GeV$ for $m_t = 171.4\ GeV$ [74]. This result is in agreement with findings in [75-77] with the delimiting scale approximately $900\ GeV$ (in agreement with the standard $1\ TeV$ scale [78]). However, more "pedantic" result [72-73], for the radiatively corrected Higgs mass in the MSSM, $m_H^2 \leq M_Z^2 + \frac{3g_W^2 M_Z^4}{16\pi^2 M_W^2}\left[\frac{2m_t^4 - m_t^2 M_Z^2}{M_Z^4} ln\left(\frac{m_T^2}{m_t^2}\right) + \frac{m_t^2}{3M_Z^2}\right]$, yields even tighter delimiting scale, $\cong 860\ GeV$ GeV, corresponding to the $m_H \leq 122.0\ GeV$ [26] for top quark mass $m_t = 175\ GeV$; which translates to $120.9\ GeV$ for $m_t = 173.1\ GeV$ and $120.0\ GeV$ for $m_t = 171.4\ GeV$. This gives a combined estimate of $m_H \leq 120.9 \pm 0.9\ GeV$. Again, a more conservative estimate may include additional $o(3\ GeV)$ uncertainty [26] for the Higgs mass, see Appendix. Also, for $m_H \cong 160.0\ GeV$ the MSSM decoupling scale becomes ~10 times larger than the HMZC scale [26] corresponding to theory that is ~100 times more finely tuned, i.e. unnatural than the SM, at low energies.

### 2.6. Two Misconceptions

The quantity $m_H^2(\Lambda^2)$, addressed above, is sometimes even described as a quantity that as a "matter of principle" [79] cannot be calculated (?). This however suggests that hierarchy/fine-tuning [40] problem cannot be quantified (i.e. ill-posed problem)? A "matter of principle", I believe, refers to the assumption that the calculation of $m_H^2(\Lambda^2)$ needs to be performed within a specific regularization scheme. Therefore if one chooses two different and supposedly equally good regularization schemes, one might get two completely different and supposedly equally good answers with unclear physical significance.

The effective Higgs mass should communicate actual measurable physical effects at large collision CM energies. Hence, it is necessary that, first, the regularization method is correctly used, second, the appropriate quantity is interpreted as the effective Higgs mass squared and third, the result has clear physical significance. As I discuss the HMZC scale is obtained by correct application of two independent, reliable methods and has clear physical significance.

Results obtained in Fig 4 are identical in both the SM dimensional $\overline{MS}$ regularization and in the Veltman's hard-cutoff method, the two most popular and most reliable approaches, to a very high precision with relatively small numerical processing error. The quantity $f = \frac{dm_H^2(\Lambda)}{d\Lambda^2}$ and all higher derivatives, i.e.

$f^{(n)} = d^n f/d(ln\Lambda^2)^n$, are completely expressible as polynomials in the SM couplings and can be entangled at the one loop level [59] with *clear physical meaning without regularization artifacts*. This is not possible at the two-loop level, but it introduces almost negligible effect [26, 33].

The interested reader should be able to easily reproduce results in Fig 2-4, in either the SM dimensional $\overline{MS}$ regularization scheme or in the Veltman's hard-cutoff method, that clearly show transition from regular to tachyonic effective Higgs particle at HMZC energies within the LHC reach. Note that $\overline{MS}$ scheme's scalar mass parameter runs only logarithmically and can be directly related to the running effective Higgs mass which runs quadratically; for details see Appendix and references therein.

Another potential misconception is the interpretation of the tachyonic Higgs and classical vacuum.

The Higgs mechanism is typically addressed in the graduate physics textbooks first by introducing a scalar field obeying the Klein-Gordon field equation with $V = \frac{m_{scalar}^2}{4}|\Phi_{scalar}|^2$, characterized by the zero scalar VEV and with a "correct" sign for the mass term, i.e. $m_{scalar}^2 > 0$. The next step is to analyze a more intricate potential, e.g. as one in Equ (1), that has a non-zero scalar VEV for $m_H^2$, $\lambda > 0$.

The standard misconception is that physics of these two theories can be directly related by renormalization; the high energy physics, with a "correct" sign, $m_{scalar}^2 > 0$, and zero VEV, can transition to the low energy physics with tachyonic $m_{scalar}^2 = -m_H^2 < 0$, corresponding to the broken electroweak symmetry.

Well, that is not exactly right! Physics of those two theories cannot be directly related by the renormalization group flow. As discussed in Appendix, it is the zero-temperature effective potential $V_{eff}$, Equ (A3), and not some particular values of the running effective parameters $m_H^2$ and $\lambda$, that defines the ground state of our today's Universe. If minima of $V_{eff}$ are away from zero, the electroweak symmetry is *broken* and non-zero VEV is characteristic of the effective theory *at all energy scales*.

If Higgs is considered to be a regular particle, i.e. with a positive mass squared, at low energies, then Higgs at energies larger than the HMZC scale, with a negative mass squared, must be considered to be a tachyon, the reason being the ground state of the world we live in. As discussed, in the zero VEV Universe, i.e. very early in the history, the tachyonic Higgs is just an ordinary, non-tachyonic particle.

### 2.7. **Meta-stable vacuum and various doom scenarios**

The assumption here is that the LHC experiment will not change the meta-stable ground state; as generally accepted, changing the meta-stable ground state, if something like that exist at all, may not be a particularly wise thing to do. The catastrophic false vacuum scenario has been addressed by Coleman [80] and Callan and Coleman [81]. Many authors also addressed the meta-stable vacuum [82-91].

In similar context, and as precautionary measure, probably the most responsible thing to do would be to have a list with various worst case theoretical scenarios, not excluded by the ultra-high energy cosmic ray arguments, see [92] for review and reference therein, a list of the corresponding experimental signatures and finally an automatic real time detection / shut down system.

I am not aware of any other more appropriate approach.

## 3. The 4D SM Higgs mass from 2D considerations

In this Section, I explain why two particular Higgs mass regions, centered at $113\ GeV$ and $143\ GeV$, may be favored on the theoretical grounds based on an analysis of the leading quantum corrections in 2D.

Both hierarchy and vacuum energy problems may be eradicated If EW symmetry breaking takes place in 2D which "embeds" the physical propagation in 4D. The leading divergences in 2D are only logarithmically divergent and therefore the large hierarchy may be easily attained. Moreover, the Higgs scalar field doesn't need to be non-zero constant in the entire 4D space hence leading to reasonable 4D space-time curvature. As addressed later on, there may be different forms of "embedding" 2D into 4D.

The various possible 2D realizations were explored in the past [93-96] and most recently in the context of "unparticle" physics [97, 98]. These are attractive research directions worth further investigations.

As well known from the lattice arguments, e.g. see [94], the non-Abelian gauge fields carry charge that causes their propagation to mimic the 1-space dimensional flux providing confinement between static charges. While QCD confinement motivated early phenomenological string theories, the above argument relaxed a need for the low energy phenomenology involving fundamental strings. In the similar spirit, the 2D considerations presented in this paper might be just an effective description, a consequence of complex dynamics of the non-Abelian gauge fields in the regular 4D space-time.

The following analysis is assumed "independent" of the physics at energies larger than the HMZC scale and it also assumes that there is no "new physics" beyond the SM at energies smaller than the HMZC scale. "Independent" in this context means described without the 2D non-SM marginal operators.

At energies smaller that the low energy HMZC scale, the electroweak $g_Y, g_W$ and top quark Yukawa $g_t$ couplings' "running" is very slow compared to the "running" of the dimensionless mass parameter $\mu = m_H^2(\Lambda)/\Lambda^2$. Also, otherwise infinite 4D scalar quadratic divergences are finite integrals in 2D. I use both of these properties to hypothesize on the physical Higgs mass.

In this context, consider loop contribution to the SM Higgs scalar propagator with the SM particles: Higgs, Z and W bosons and top quark, in the loops. The 2D loop calculations are addressed in Appendix.

I split the top loop to two, $x$ and $y$, contributions. Part proportional to $x$ exactly cancels the Higgs and gauge boson loops while piece proportional to $y$ equals radiatively generated Higgs mass, see Fig 5.

The calculation is self consistent as the Higgs mass in the Higgs loop propagator (within piece proportional to $x$) is identical to radiatively generated Higgs mass (within piece proportional to $y$).

Therefore, this is analogous to absence of the bare mass term; Higgs mass is explicitly and self-consistently generated by top loop. Cancelation of leading "divergences" may point out to important relationships between the physical quantities. Here, I assume that $x$ and $y$ add to $z$ equals 1 or $\frac{2m_t^2}{v_{EW}^2}$.

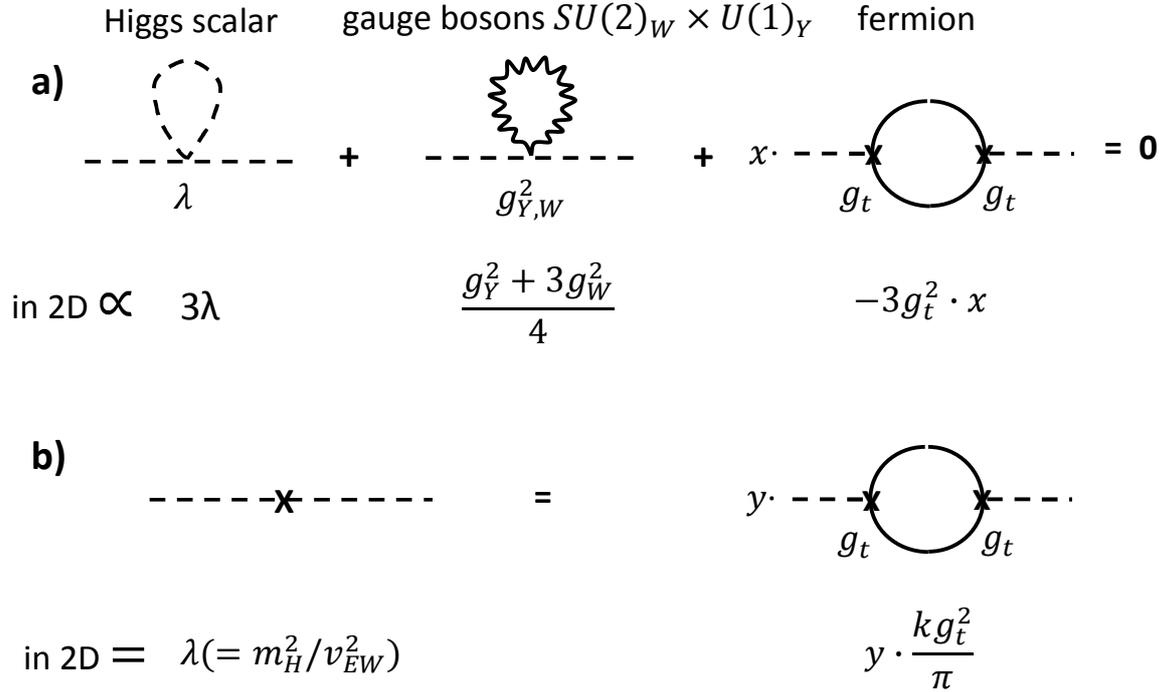

**Figure 5** The 2D **SM a) cancelation of leading quantum corrections and b) radiatively generated Higgs mass.**

Top loop in 2D is exactly solved as

$$top\ loop = \frac{kg_t^2}{\pi}\ .\qquad(2)$$

This loop calculation is compatible with the famous Schwinger result for 2D massive QED with photon vacuum polarization generating mass $\frac{e}{\sqrt{\pi}}$ for the gauge boson [93]; see also [94] and references therein. The above loop may be obtained from the Schwinger model by anti-commuting gamma matrices within the trace of the 2D transversal gauge boson propagator to show that momentum-independent part in 2D Schwinger model equals the scalar propagator times the metric tensor.

Clearly, physics presented here is very different from the Schwinger model and correspond to the "full" SM calculation of the radiative corrections in 2D in leading order and applied to the scalar sector.

To make connection with 4D, factor of 2 is expected here, due to the scalar nature of propagator, as top quark spins from top loop may point inward or outward (when 4D spin is also considered). Hence, I leave an explicit dependence on the relevant phase space parameterized with $k = 1$ (2). The obtained system has three unknowns $x, y$ and $\lambda$, and three equations,

$$3\lambda + \frac{g_Y^2+3g_W^2}{4} = 3g_t^2 x\ ,\quad x+y=z\quad \text{and}\quad \lambda = \frac{kg_t^2}{\pi}y\ ,\qquad(3\ a,b,c)$$

leading to an unique solution
14

$$3\lambda + \frac{g_Y^2+3g_W^2}{4} = 3g_t^2\left(z - \frac{\lambda\pi}{kg_t^2}\right) \rightarrow \sqrt{\lambda} = \sqrt{\frac{g_t^2 z - \frac{g_Y^2+3g_W^2}{12}}{1+\frac{\pi}{k}}}, \tag{4}$$

or in terms of the physical Higgs mass

$$m_H = \sqrt{\frac{6m_t^2 z - M_Z^2 - 2M_W^2}{3\left(1+\frac{\pi}{k}\right)}}. \tag{5}$$

In the range of the top quark mass, $m_t = 170 - 175\ GeV$, the above result varies as

$$m_H = \begin{cases} 110.7 - 114.4\ (107.7 - 115.0)\ GeV\ for\ z = 1\ or\ (\frac{2m_t^2}{v_{EW}^2}),\ k = 1 \\ 140.5 - 145.2\ (136.7 - 146.0)\ GeV\ for\ z = 1\ or\ (\frac{2m_t^2}{v_{EW}^2}),\ k = 2 \end{cases}. \tag{6}$$

For the *current* world average top quark mass, $m_t = 173.1 \pm 1.3\ GeV$ [69], I obtain

$$m_H = \begin{cases} 113.0\ (112.3) GeV\ for\ z = 1\ or\ \left(\frac{2m_t^2}{v_{EW}^2}\right),\ k = 1 \rightarrow y = \mathbf{0.669} \\ 143.4\ (142.5) GeV\ for\ z = 1 or\ \left(\frac{2m_t^2}{v_{EW}^2}\right),\ k = 2 \rightarrow y = \mathbf{0.539} \end{cases}. \tag{7a}$$

Whereas for the several years older value, $m_t = 171.4 \pm 2.1\ GeV$ [74], I obtain

$$m_H = \begin{cases} 111.9\ (110.0) GeV\ for\ z = 1\ or\ \left(\frac{2m_t^2}{v_{EW}^2}\right),\ k = 1 \rightarrow y = \mathbf{0.669} \\ 142.0\ (139.6) GeV\ for\ z = 1 or\ \left(\frac{2m_t^2}{v_{EW}^2}\right),\ k = 2 \rightarrow y = \mathbf{0.539} \end{cases}. \tag{7b}$$

This result should be corrected by the effects of the "running" electroweak $g_Y, g_W$ and top quark Yukawa $g_t$ couplings between $m_H$ and $\Lambda_{HMZC} \cong 1\ TeV$. This will be addressed in detail elsewhere. A preliminary analysis suggests that this correction is in the order of magnitude of maximally a few percents.

There is an additional uncertainty for the physical Higgs mass, due to the finite cut-off. The top loop is exactly solved in Equ (2) with an infinite cut-off. For example, for the k=1 branch, the finite cut-off scale equals the HZMC scale $\sim 10^{2.9} \cong 800\ GeV$ [26] at which the effective Higgs mass is zero for the physical Higgs mass $m_H = 115.0\ GeV$. Therefore, the cut-off effects are $\leq m_H^2/\Lambda_{cut-off}^2$ and may introduce additional uncertainty only on the order of $\frac{3}{8\pi^2} 2\% \cong 0.01\%$ for the physical Higgs mass obtained above.

Clearly, the "new physics" marginal operators may also include an additional uncertainty.

Hence, the k=1 mass branch embraces the late LEP Higgs signal candidate [23-24, 44-46].

If the SM smoothly expands and provides complete description above the HZMC scale, within the Type II transition (which may correspond to either the first or the second order EW phase transition), there is another cut-off scale anywhere between ~1 TeV and ~500 TeV due to the vacuum stability limit [26].

The k=2 branch has the HZMC scale at $\sim 10^{3.03} \cong 1100\ GeV$ for the physical Higgs mass $m_H = 142.0\ GeV$. It is worth noting that the k=2 mass branch masses are in the center of the Higgs mass range favored by the combined electroweak precision data global fit and direct searches [46-47], see Section 2.5. And the combined CDF and DØ analysis [49] observed the $1\sigma$ excess for $132\ GeV < m_H < 143\ GeV$ Higgs range. If the SM smoothly expands and provide complete description above the HZMC scale, within the Type II transition (which may correspond to either the first or the second order EW phase transition), there is, interestingly, no stability and perturbativity constraining scales below the Planck mass. Furthermore, both SM dimensionless parameters, $\lambda$ and $= m_H^2(\Lambda)/\Lambda^2$, are independently approaching zero in the close vicinity of the Planck mass energy scale. Their deviations from zero are smallest for $m_H = 137.6\ GeV$ [33] and $m_t = 175\ GeV$, see Fig 7; this value is slightly shifted to $m_H = 138.1\ GeV$ due to corrections associated with the current top quark mass world average, $m_t = 173.1 \pm 1.3\ GeV$.

### 3.1. "Embedding" 2D dynamics within 4D?

Here, I hypothesize, without proof, that standard 4D EWSB might be only an useful description that corresponds to dynamical, effective 2D description as this can solve hierarchy and vacuum energy problems that particle physics faces today. In 2D leading quantum corrections are only logarithmically divergent and scalar VEV (or ground state expectation value of the scalar field) may be confined only to propagator 2D space associated with propagating particles in 4D; i.e. equivalent to compactifying the Higgs *ether* from entirety of 4D to just a small subset of space and hence leading to realistic space-time curvature. I addressed these ideas in the past [33].

A complete removal of the Higgs *ether* would be dynamical and *local* symmetry breaking, when described in terms of the more fundamental ("new") physics at short distances, i.e. energies above the HMZC scale. Hence, with effective scalar field being non-zero, and approximately constant, most likely only in the close vicinity of the 4D particles at large distances, i.e. energies below the HMZC scale.

Clearly, there may be different forms of "embedding" 2D into 4D. And the lower dimensional 2D space may or may not be orthogonal to two polarizations of the 4D massless gauge bosons.

"Embeding", for example, can be to the extent that (1) 4D electroweak symmetry breaking is governed by 2D electroweak symmetry breaking and 4D couplings or (2) that 4D theory is effective theory completely described by 2D theory, where dimensionality of space-time enters less as a premise and more as a consequence of the fundamental 2D theory. The later idea is briefly addressed in Section 3.2.

### 3.2. Radical possibility (though maybe too radical)

If 2D fermions have internal degrees of freedom that in combination with external degrees of freedom transform in the "right" way under the 4D Poincaré group, i.e. the inhomogeneous Lorentz group, then there is no *apriori* reason why such a model cannot be interpreted as 4D. For example, the internal degrees of freedom would contribute to the 4D 4-momentum which may or may not be "on shell".

To illustrate this idea a bit further I shall sketch the simplest possible example. Consider that 2D fermions are described by two "flavors", A and B. Furthermore, imagine that the 4D fermion (dimension 3/2) may be constructed out of the 4D scalar field (dimension 1) and a 2D fermion (dimension ½). Moreover, assume that the 4D scalar may be interpreted as a 2D condensate composed of left and right moving A or B or their linear combination. On these lines, the 4D vector boson, e.g. the transversal spin 1 component, may be interpreted as a linear combination of two 2D fermions moving in the same direction. And the 2D vertex may appear as 4D if there is additional phase space attached to the interacting 2D particle. For example that phase space may be a consequence of one $SU(2)_A$ or $SU(2)_B$ symmetry that may be related to $SU(2)_L$ or $SU(2)_R$. After all, the Lorentz group is related to $SL(2,C)$, which is $SU(2)_L \times SU(2)_R$. Finally, Higgs mechanism may be confined to 2D and described by the non-zero condensate VEV.

Why is this idea extremely fascinating? Well, apart from almost disturbing notion that our ordinary 4D space-time might be compactly "written" in 2D, this concept could have many interesting implications on physics; it may clearly render vacuum energy and hierarchy problems non-existent and it could have a deep impact on the current notion of gravity [37-38].

As discussed in Section 5, condensation is likely closely related to the interplay of "new physics", i.e. various marginal operators and 4 fermion interactions which are *renormalizable* interactions in 2D, with QCD. Furthermore, see Section 7.1, one may naturally expect a bound state built out of three fermions to "balance" electroweak gauge bosons' and Higgs loops at the energies smaller than the electroweak symmetry breaking scale while potentially *exactly* removing the tachyon solution at all larger energies.

## 4. Top condensation and "new physics" model building: brief overview

Here, I briefly introduce the dynamical mass generation involving top condensation; see for example [99-100] for a more complete review.

The concept of the dynamical mass generation and spontaneous symmetry breaking, potentially explaining the electroweak symmetry breaking in the particle physics, was built upon the pioneering work on the *"microscopic" theory of superconductivity* by Bardeen, Cooper and Schrieffer [101]. This concept was revisited, further advanced, and introduced in high-energy particle physics by Nambu and Jona-Lasinio [102-104] as the NJL model. At the similar time, the methods of theory of superconductivity and their application to the mass generation in particle physics were addressed by Vaks and Larkin [105]. The NJL model was applied by Hill [32], Miransky et al. [106] and Bardeen et al. [107]. The assumption there is that strong effective 4-fermion interactions may trigger the top quark condensation, hence, introducing a composite effective Higgs scalar that has exactly the right quantum numbers to break the EW symmetry in a dynamical manner. In a difference to the Technicolor [108-109, 40] models where new particles, technifermions condense, it is the SM top quark degrees of freedom here that are anticipated to be responsible for the EW symmetry breaking.

It has also been shown that there is no fundamental theoretical obstacle that would prohibit the composite effective Higgs particle to completely mimic the SM "fundamental" Higgs particle at low energies [110].

As emphasized by Nambu [111] the Higgs mass is determined as $m_H = 2m_t \cong 350\,GeV$ within the gauged Nambu-Jona-Lasinio mechanism when applied to the SM and implemented in the fermion loop approximation [107]. The minimal model, which attempted to incorporate the "full" SM, as an effective low energy theory, was proposed by Bardeen et al. [107]. However, this model predicted too large top quark mass in the close vicinity of the SM renormalization group, QCD driven, top quark infrared fixed point $\sim 230\,GeV$ [32,107] and large Higgs mass $\sim 260\,GeV$. Somewhat lighter but still heavy Higgs, $150 < m_H < 250\,GeV$, has been obtained by two independent studies by Cvetic [112] and Hambye [113].

Hence, it was observed that smaller top quark masses generally could not provide enough electroweak breaking VEV to create the appropriate masses for the W and Z bosons. That can be also verified, for example, with the Pagels Stokar relationship [114] or with gap equations in the gauged NJL model [115,116] where one obtains the $f_t$, i.e. the top analog of the pion decay constant, $f_\pi$, that seems to be too small. The observed mismatch between the SM fermion and boson masses motivated the Topcolor model [117], with a new strong interactions singling out the top quark, as well as the class of models which combined Topcolor with Technicolor within a model building effort termed the Topcolor-assisted-Technicolor, TC^2 [118-126]. Another related approach, the Topcolor Seesaw [127-129] models applied a seesaw type of mixing among the "new" fermions, either the weak singlets [127,128] or the weak doublets [129], with a goal to lower the dynamically generated quark mass. Some of the top condensate model building efforts also incorporated physics of extra dimensions [130,131].

Without experimental data that may directly confirm or reject particular theoretical concepts, the majority of above models should be considered as quite attractive and viable possibilities, though most likely highly incomplete. For example, they may require additional model structure to generate the realistic particle mass spectrum. On those line, experience with the Extended Technicolor [132,133] was that a lot of thought has to be given to the flavor changing neutral currents [132,133], unwanted contributions to $R_b$ [134], excessive isospin violation [135] etc.

## 5. Top condensation consistent with gluon and Higgs scalar mediation?

Here, I investigate whether the SM $\bar{t}t$ channel is repulsive or attractive as a necessary SM condition for, as anticipated here, an almost-loose bound state.

I assume that there is an underlying dynamics that correlates $\bar{t}t$ values and orientations among different space-time points as well as among different momentum eigenstates. However, I assume that this underlying dynamics is completely (or almost completely) expressible with the SM degrees of freedom, thought of here to represent the low energy effective theory in Wilson's approach [39].

Hence, I investigate whether the SM $\bar{t}t$ channel is repulsive or attractive as a necessary condition for an almost-loose bound state. I postpone more advanced analysis to Sections 6 -7. Physics presented here is different from analysis linking the top quark mass with the QCD driven infrared fixed point [32,107].

Consider the $\bar{t}t$ scattering in the Euclidean space and ignore chiralities of the incoming and outgoing particles while assuming that left and right handed tops are equally represented within particle and antiparticle solution. Main interaction channels at tree level are gluon and Higgs exchange.

The weak interactions are absent as interacting particles have opposite chiralities and the hypercharge interactions are zero due to the equal sharing conjecture introduced above.

I now assume that strong QCD interactions proportional to $-g_{QCD}^2 T_{aij}T_{akl}$, where $a = 1 \ldots 8$, $i,j = 1,2,3$ and summation over repeated indices is implied, are exactly *balanced* with the Yukawa forces due to the virtual Higgs particle exchange proportional to $g_t^2$ as condition for the loose bound state; see Fig 6. Hence, the back of the envelope calculation at tree level suggests

$$2\frac{1}{2}\frac{2}{3}g_{QCD}^2 = g_t^2 \quad or \quad \alpha_S = \frac{3}{2}\frac{g_t^2}{4\pi} \tag{8}$$

where I used $(T^a)_{ij}(T^a)_{kl} = \frac{1}{2}\left(\delta_{il}\delta_{kj} - \frac{1}{N}\delta_{ij}\delta_{kl}\right)$ for $SU(N)$ groups, see for example [136], where $N = 3$. For the *colorless* composite Higgs the expression in bracket equals $\frac{2}{3}$, and additional factor of 2 in Equ (8) corresponds to two transversal gluon polarizations. **QED**

Result in Equ (8) is in an excellent agreement with the standard estimate of the strong running coupling constant [46, 133]. The above result predicts $\alpha_s = 0.1181 \pm 0.0018$ for the world average top quark mass $m_t = 173.1 \pm 1.3 \, GeV$ [69] where uncertainty is therefore solely due to the top quark mass uncertainty. This can be compared with the current world average value $\alpha_s \cong 0.1184 \pm 0.0007$ at $s = M_Z^2$ [46, 137].

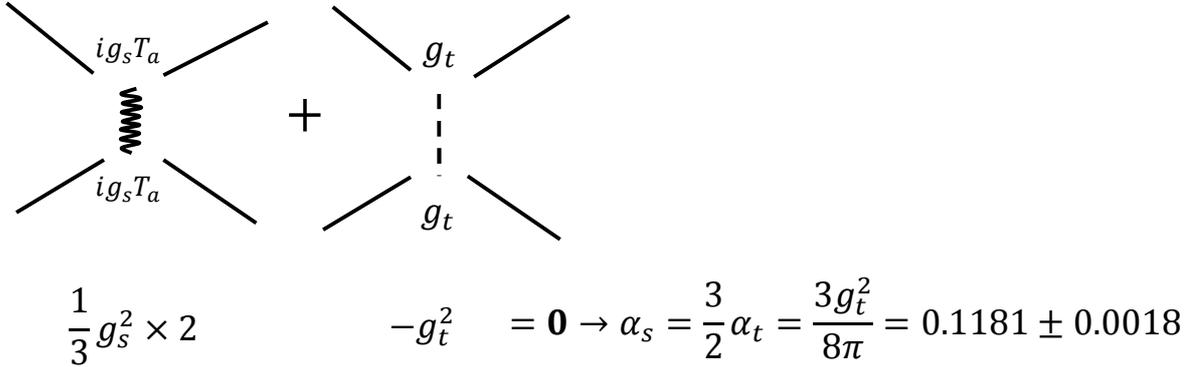

**Figure 6** Finely balanced interplay between the QCD gluon and Higgs scalar mediated top anti-top interactions.

Even if the equal distribution assumption is ignored and hypercharge interactions are taken into account that would change the above result only on the order $\frac{3}{2}Y_L Y_R \frac{g_Y^2}{g_t^2} = \frac{3}{2}\frac{1}{6}\frac{2}{3}\frac{g_Y^2}{g_t^2} \sim 2\%$. And if this difference is to be accounted for example by difference between $\alpha_s(M_Z^2)$ and $\alpha_s(m_H^2)$, then even in the next to leading order correction it appears that lighter Higgs, e.g. in the vicinity of $m_H \sim 115 \, GeV$ with $\alpha_s(m_H^2 = 115 \, GeV) \sim 0.105$, would be more preferred than the heavier one, $m_H \sim 145 \, GeV$. The lighter Higgs tends to be more compatible with the Type I transition, see Section 7. I postpone making a definite conclusion until the leading corrections are properly addressed elsewhere.

If above observations are correct then interplay between logarithmically running top Yukawa coupling constant and logarithmically running strong coupling constants may indeed define the low energy HZMC scale, at which the top condensate forms and breaks the electroweak symmetry within Type II (I) transition as addressed in Section 6 (7). Anyhow, this is good news as this symmetry breaking principle, or maybe only contributing principle, may span vast energies in a natural fashion.

Finally this interplay may be completely *local* and bound to the finite volumes surrounding propagating particles [33]. For the loose bound state it takes small or no energy to locally order the background condensate field. If this is a local, dynamical process it may clearly resolves the *vacuum energy problem*.

**Q:** How the light Higgs ($m_H < 182\ GeV$) can be represented as a loose bound state of two top quarks $m_t \cong 173.1\ GeV$ when mass of two top quarks is $\cong 346.2\ GeV$?

**A:** As I argue in the Section 7 the top condensate in 4D probably includes 3 effective scalar fields with mass $m_H \cong \frac{2}{3} m_t \cong 115.4\ GeV$ closely related to 2D k=1 case dynamics analyzed in Section 3. Also, as well known from the lattice arguments, the non-Abelian gauge field carries charge that cause its propagation to mimic 1 space dimensional flux providing confinement between static charges [94].

**Q:** Could solution for hierarchy and vacuum energy problems come from 2D theory?

**A:** Maybe. The importance of 2D theory was heavily explored in the past [93-96] and most recently in the context of *unparticle* physics [97, 98]. It surely seems to be attractive option worth investigation.

## 6. Higgs at very high energies and second order phase transition

In this section I hypothesize on physics that may be responsible for the $142 \pm 6\ GeV$ masses, as obtained in the Section 2, or k=2 Higgs mass branch $137 - 146\ GeV$, as obtained in Section 3. I show that the heavier Higgs solution is likely associated with the "desert scenario", or the "very long lived" SM, within the Type II transition. Here, the effective theory does not abruptly change the parameters and degrees of freedom across the low energy HMZC scale affiliated with the Type II transitions.

As I explain next, it is possible that the SM with composite Higgs in 4D may span vast energy scales. The hierarchy problem, in this case, seems to be a rather benign problem for $m_H^2(\Lambda^2) < 0$.

### *6.1. Two special solutions affiliated with the Planck mass scale*

Within the entire high-energy SM effective theory "spectrum" there is a single region where both dimensionless parameters $\mu \equiv m_H^2(\Lambda^2)/\Lambda^2$ and $\lambda$ almost coincide with zero value, see Fig 7. Interestingly enough this is in the vicinity of the Planck mass, obtained as a consequence and not as a premise. This solution to conjecture that minimize the parameters of the Higgs potential is obtained for the physical Higgs mass centered at $m_H = 137.6\ GeV$ [33]; this result, however, is slightly shifted to $m_H = 138.1\ GeV$ to accommodate for the current top quark mass world average. This solution overlaps with both the $142 \pm 6\ GeV$ Higgs mass range, obtained in Section 2, and the k=2 mass branch, obtained in Section3.

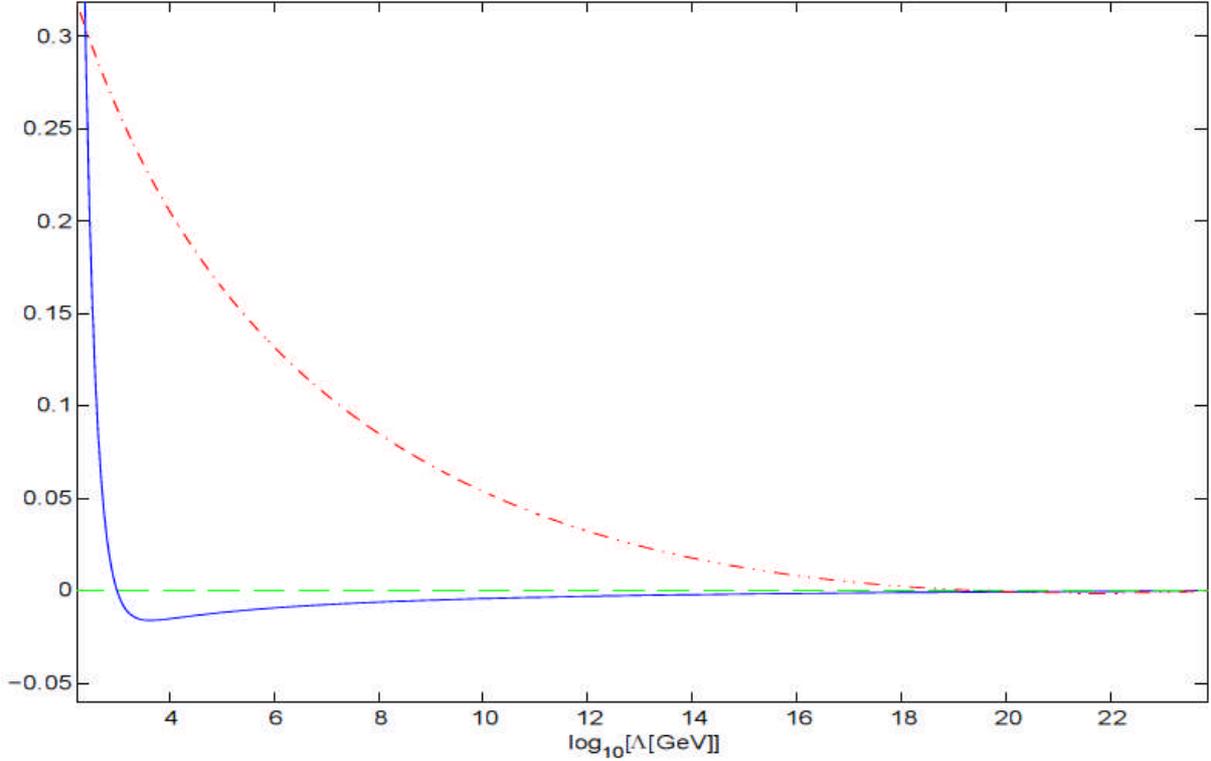

**Figure 7** The running of the SM Higgs quartic coupling $\lambda$ and dimensionless parameter $\mu \equiv m_H^2(\Lambda^2)/\Lambda^2$, i.e. rescaled effective mass squared, for the SM Higgs candidate, physical mass $m_H \cong 138\ GeV$, for the physics originating at the very high energy corresponding to roughly the Planck mass [33].

While it is traditionally anticipated that $\mu$ should run quadratically, the actual SM $\mu$ for the $m_H = 138\ GeV$ solution, Fig 7, runs logarithmically at energies larger than the HMZC scale. Moreover, already based on visual inspection, it appears that the SM dimensionless parameters, $\lambda$ and $\mu$, are correlated.

The parameter $\mu$ at the Planck scale is exactly equal zero for $m_H \approx 146.5\ GeV$; this is the Planck mass adaptation of the Coleman-Weinberg (CW) conjecture where bare mass (though and not $m_H(M_{Pl})$ as here) is zero and the electroweak breaking is governed by the quantum corrections [34]; see[26, 33].

### *6.2. Composite Higgs from very high energies*

As I show next, the *slow* "running" or better "walking" should be expected from the composite Higgs for positive $\lambda$ and negative $m_H^2(\Lambda^2)$. If one ignores the higher order corrections the running has simple functional dependence supporting a "long lived" solution

$$\mu \propto -\sqrt{\lambda}\ . \tag{9}$$

Main idea here is that Higgs field might be an "almost" fundamental field generated in the proximity of the Planck scale or some other high energy scale $\Lambda_{high}$, i.e. effective field composed out of fermion degrees of freedom with assumed zero potential energy density. Composite Higgs in the context of top condensate electroweak breaking has been addressed in the past [107,111]. Also, it has been shown that theory with composite Higgs may indeed mimic the minimal SM with fundamental Higgs scalar field at low energies [110]. However, as I show here, there are still a few constraining features which if not satisfied may tell apart elemental from the composite Higgs particle.

Just beneath the high energy scale $\Lambda_{high}$ the Higgs scalar acquires correct couplings to gauge boson fields through field renormalization and top Yukawa coupling renormalization, see Fig 8, such that

$$-I(\Lambda)g_{t*}^2 Z_\Phi^{-1} = 1 \qquad (10)$$

where $g_{t*}$ is the bare top Yukawa coupling and top loop integral equals $-I(\Lambda)g^{\mu\nu}$ where $\Lambda \leq \Lambda_{High}$.

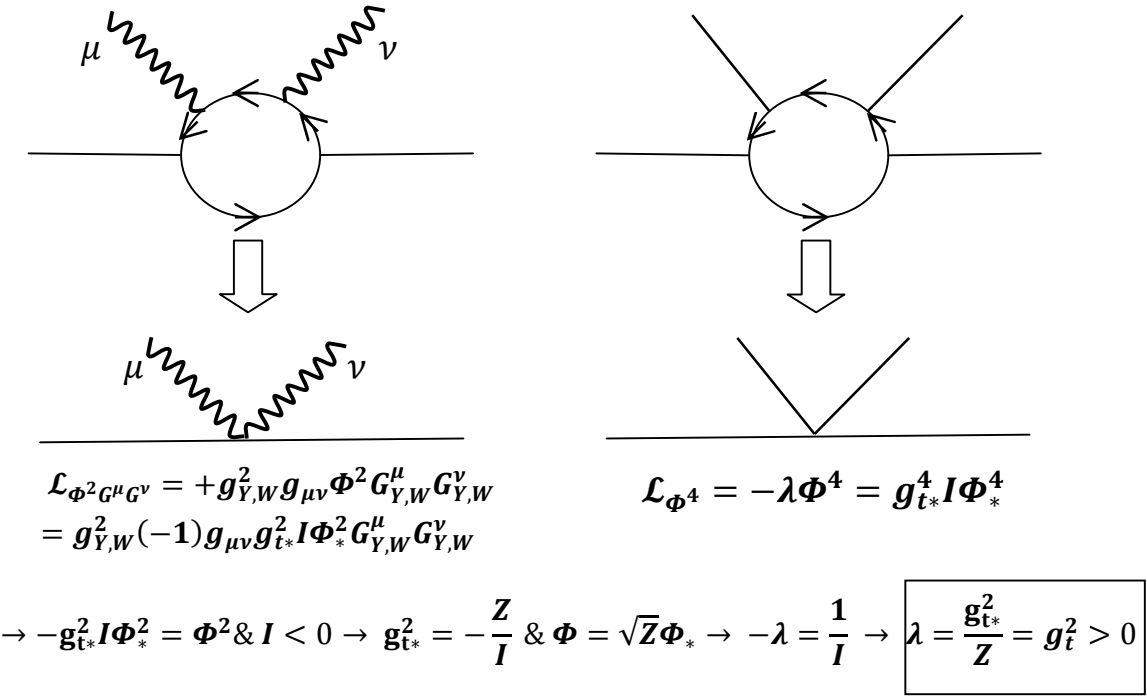

**Figure 8** Top loop induced generation of the Higgs scalar electroweak couplings and quartic coupling.

Similarly, the Higgs scalar acquires quartic coupling subject to

$$I(\Lambda)g_{t*}^4 Z_\Phi^{-2} = -\lambda(\Lambda) \qquad (11)$$

where top loop integral equals $I(\Lambda)$ proportional to $ln(\Lambda^2)$ in 4D.

Interested reader should be able to prove that $I(\Lambda)$ is common for both loops.

By dividing Equ (11) with square of Equ (10) one obtains a *positive* effective quartic coupling $\lambda(\Lambda)$, i.e.

$$\lambda = -\frac{1}{I(\Lambda)} = g_{t*}^2 Z_\Phi^{-1} = g_t^2(\Lambda) > 0 \; . \qquad (12)$$

The above solution should go to zero with $\Lambda \to \Lambda_{High}$ and should have a zero potential energy density at $\Lambda_{high}$. Below this scale coupling runs "logarithmically" and is renormalized according to the SM renormalization flow. However, being the characteristic of composite theory it nonetheless conspires to mainly reproduce the leading order term Equ (12). It would be interesting to investigate which solution within the $k = 2$ Higgs mass branch expresses this property the most; I will address that elsewhere.

Consider next the composite dimensionless Higgs mass squared $\mu$ which is radiatively generated through top loop, see Fig 9, in 4D. In difference to Sec 3 solution this solution should have a zero value at the high energy scale and subsequently, just beneath that scale, smoothly gains its *negative*, therefore tachyonic, effective mass squared, as a consequence of the minus sign associated with the fermion loop.

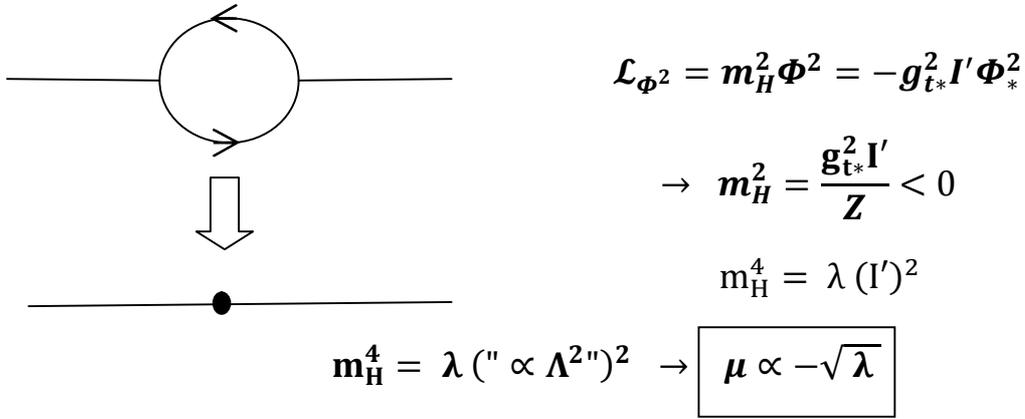

**Figure 9** Radiative generation of the Higgs scalar mass from top loop.

According to Wigner [138], the space-like, negative mass squared, particles have non-compact little groups; their spin is not described by rotation group $SU(2)$ and in difference to massless particles their "spin" may be continuous parameter, see [139]. This might be related to the extended Higgs sectors I introduce in Sections 6-7. But, the general concept of Higgs tachyon solution requires better understanding. Hence, I hope, this paper will motivate more focused research effort in that direction.

The top loop $I'(\Lambda)$ in 4D is negative and proportional to quadratic term $\Lambda^2$

$$I'(\Lambda^2) \sim \Lambda^2 \;\; \to \;\; \mu(\Lambda^2) \propto -\sqrt{\lambda(\Lambda^2)}. \tag{13}$$

Therefore, as anticipated in Equ (9), the dimensionless mass squared, $\mu$, in the leading order is expected to be proportional to the square root of the scalar quartic coupling which runs only "logarithmically"; meaning no traditional hierarchy problem for the minimal SM with composite Higgs in 4D!

After many orders of magnitude the higher order corrections should overcome the composite Higgs mass leading order, and $\mu$ should finally reach the zero value corresponding to the low energy HMZC scale – condition for the electroweak phase transition. See Section 5 for interpretation of the interplay between $g_t$, $g_{QCD}$ and their effect on $\mu$. Finally, shortly beneath the HMZC scale, the renormalization flow drives $\mu$ to the intersection with $\lambda$ at which point the smooth second Type II transition is completed with the correct value of the Higgs' VEV. Short running below the HMZC scale is a natural consequence of, now, positive Higgs mass squared and the renormalization flow at low energies.

Therefore, $m_H = 138.1 \pm 1.8 \, GeV$ might be a good candidate for the composite Higgs mass, high energy fundamental scale placed in the vicinity of the Planck scale and the electroweak symmetry breaking scale $\cong 1050 \, GeV$ within the Type II transition.

It seems to me that effort to quantify deviation from Equ (13) across $142 \pm 5 \, GeV$ mass range, and across energies smaller than the Planck mass, can be worthwhile; I will address that elsewhere.

As shown in Section 5, the SM scale at which tops condense may coincide with the low energy HMZC scale. This may be interpreted with a dual model in which, instead of an almost "fundamental" Higgs particle, one considers the original high energy model structure. The unknown elements of that model, e.g. marginal and 4-fermion interactions, extra or less dimensions etc., that single out the dominant top quark, hence, conspires, in a natural fashion and in accord with the QCD, to create the EW symmetry breaking condensate at just the "right" low energy scale corresponding to the low energy HMZC scale.

In a summary, the $142 \pm 5 \, GeV$ solutions, supporting the composite Higgs originating at very high energy scales, and mimicking the minimal SM Higgs at lower energies, are favored for the following reasons:

1) These are the values in the center of the currently favored SM Higgs mass range as obtained by the combined electroweak precision data global fit and Higgs direct searches; see Section 2.5. And the combined CDF and DØ analysis [49] observed the $1\sigma$ excess for $132 \, GeV < m_H < 143 \, GeV$ Higgs range.

2) The leading divergences cancel out with the consistent value of radiatively generated Higgs mass; see k=2 branch solutions, presented in Section 3.

3) The Higgs and gluon mediated top interactions might satisfy condition for the $\bar{t}t$ loose bound state at low energies; see Section 5.

4) There are no vacuum stability and perturbativity constraints at energies smaller than the Planck mass and there is a single HMZC scale (i.e. no multiple HMZC scales) in the same energy range; see Section 2.

5) Condition $\mu, \lambda = 0$ at $\Lambda_{high}$ is *most closely* satisfied for $\Lambda_{high} \sim M_{Planck}$ and $m_H = 138.1 \, GeV$

6) Condition $\mu(\Lambda) < 0$, $\lambda(\Lambda) > 0$ is satisfied for $\Lambda < \Lambda_{high}$ in the vicinity of $\Lambda_{high} \sim M_{Planck}$.

7) Hierarchy problem for $m_H^2(\Lambda^2) < 0$, $\lambda(\Lambda) > 0$ and composite $\mu(\Lambda^2) \propto -\sqrt{\lambda(\Lambda^2)}$ appears to be benign.

8) Finally there is a way, that I show next, to directly match the heavy 4D Higgs with light k=1 2D branch solutions, presented in Section 3, implications of which are discussed in Section 7.

### 6.3. Model $3^{-1/2}$ within the Type II transition

Consider 4D Lagrangian density

$$\mathcal{L} = \sum_k \left\{ [D_\mu \Phi_k]^+ [D^\mu \Phi_k] + \sum_i g_i \bar{\Psi}_i \Phi_k \Psi_i \right\} - \frac{m_H^2}{4} \sum_k \Phi_k^+ \Phi_k + \frac{\lambda}{8} \left( \sum_k \Phi_k^+ \Phi_k \right)^2 \quad (14)$$

where $k = 1 \ldots 3$ and $i$ counts fermions. Assume that each scalar field $\Phi$ develops a non-zero vacuum expectation value equal to $\langle|\Phi|\rangle = \langle|\Phi_{SM}|\rangle/\sqrt{3}$. Hence, the fermion Yukawa coupling is $g_i = g_{SMi}/\sqrt{3}$ and the scalar - gauge bosons coupling is the same as in the SM.

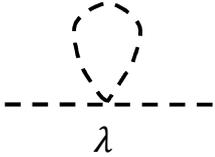

**Figure 10** Cancelation of leading quantum corrections for Higgs scalar propagator in 4D Model $3^{-1/2}$.

The conditions for cancellation of quadratic divergences, see Appendix, in the scalar propagators are

$$\lambda + \frac{g_Y^2 + 3g_W^2}{4} - \frac{2g_t^2}{3} = 0 \quad \rightarrow \quad \sqrt{\lambda} = \sqrt{\frac{2g_t^2}{3} - \frac{g_Y^2 + 3g_W^2}{4}}. \quad (15)$$

Or in terms of observed masses, the Higgs mass is

$$m_H = \sqrt{\frac{4}{3} m_t^2 - M_Z^2 - 2 M_W^2} = 138.3 \text{ GeV for } m_t = 174 \text{ GeV}. \quad (16)$$

This result was obtained in [33] and it is in good agreement with the Higgs mass obtained in the previous subsection. In the range $m_t = 173.1 \pm 1.3$ GeV one obtains a slightly smaller value $m_H = 136.8 \pm 2.2$ GeV.

If one now tries to reproduce the result on radiatively generated Higgs mass one obtains

$$\lambda v_{EW}^2 = \left(1 - \frac{2}{3}\right) \cdot \text{top loop} \quad \text{or top loop} = 3\lambda v_{EW}^2 = 3 \frac{m_H^2}{v_{ew}^2} v_{EW}^2 = 0.93 v_{EW}^2, \quad (17)$$

or prediction for the "top loop" in 4D. It is interesting that for the central value within the k=2 branch

$$\text{top loop} = 1 \cdot v_{EW}^2 \quad \text{for} \quad m_H = 142.5 \text{ GeV}. \quad (18)$$

As discussed in Section 5, an interesting interplay between QCD and Higgs scalar exchange may be potential reason for just the "right" HMZC scale, i.e. the $\sim 1$ TeV electroweak symmetry breaking scale.

This model in notation of Section 3 has $x_1 = x(k=1) = 0.331$ and $y_1 = y(k=1) = \mathbf{0.669}$ that are characteristics of the k=1 branch; however, in 4D, interestingly this corresponds to the k=2 Higgs mass.

For the current world top mass average, $m_t = 173.1 \, GeV$, scaling between Equ (15) in 4D and Equ (3) for 2D k=1 case, see Section 3, corresponds to $\cong \mathbf{1.4676}/3$ and $\cong 2/3$ for scalar and top loop respectively.

The "curious" number connecting 2D (k=1) and 4D theories can be expressed as $\frac{\lambda_{4D}}{\lambda_{2D}} = 1.4676 = \left[\frac{2}{3} - 3\left(x_1 - \frac{y_1}{\pi}\right)\right] / \left(\frac{y_1}{\pi}\right)$. Interested reader should be able to easily derive this relationship.

I use this scaling in Section 7, to show how to conserve theory structure, in particular cancelation of leading divergences, across space-times with different dimensions, below and above the HMZC scale.

### 6.4. *Model $\pi^{-1/2}$ in 2D*

As next excursion into "new physics" consider a model embedded in 2D with Lagrangian density

$$\mathcal{L}_0 = \int_0^\pi d\theta \left\{ [D_\mu \Phi(\theta)]^+ [D^\mu \Phi(\theta)] + \sum_i g_i \bar{\Psi}_i \Phi(\theta) \Psi_i \right\}, \tag{19}$$

whereas the Lagrangian density describing fermions' interaction with gauge bosons is

$$\mathcal{L}_1 = \sum_i \bar{\Psi}_i D_\mu \gamma^\mu \Psi_i, \tag{20}$$

where $D_\mu = \partial_\mu + ig_Y Y B_\mu + ig_W \vec{T} \vec{W}_\mu$. The scalar self-interaction is of the form

$$\mathcal{L}_2 = \int_0^\pi d\theta \left\{ -\frac{m_H^2}{4} |\Phi(\theta)|^2 + \frac{\lambda}{8} |\Phi(\theta)|^2 \int_0^\pi d\theta' |\Phi(\theta')|^2 \right\} \tag{21}$$

Each scalar field $\Phi(\theta)$ obtains a non-zero VEV equal to $\langle |\Phi(\theta)| \rangle = \langle |\Phi_{SM}| \rangle / \sqrt{\pi}$ and

$$\langle |\Phi|^2 \rangle = \int_0^\pi d\theta \, \langle \Phi(\theta) \rangle^2 = \int_0^\pi d\theta \left( \frac{\langle \Phi_{SM} \rangle}{\sqrt{\pi}} \right)^2 = \langle \Phi_{SM} \rangle^2 \tag{22}$$

provides the appropriate masses for the Z and W gauge bosons. Similarly, the fermion Yukawa coupling must be $g_i = g_{SM}/\sqrt{\pi}$ and the scalar - gauge bosons coupling is the same as in the SM.

Clearly, the idea is that maybe

$$\langle Re(\Phi) \rangle = \int_0^\pi d\theta \, \langle \Phi(\theta) \rangle \cos\theta = \int_0^\pi d\theta \, \frac{\langle \Phi_{SM} \rangle}{\sqrt{\pi}} \cos\theta = 0 \tag{23}$$

may provide alternative explanation for the vacuum energy problem.

The condition for cancellation of quadratic divergences for each scalar field $\Phi(\theta)$ is shown in Fig 11.

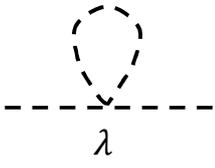

Figure 11 Cancelation of leading quantum corrections for Higgs scalar propagator in 2D Model $\pi^{-1/2}$.

The above result suggests that

$$\sqrt{\lambda} = \sqrt{\frac{1}{3}\left(\frac{3}{\pi}g_t^2 - \frac{g_Y^2 + 3g_W^2}{4}\right)} \rightarrow m_H = \sqrt{\frac{1}{3}\left(\frac{6}{\pi}m_t^2 - M_Z^2 - 2M_W^2\right)}, \text{i.e.} \quad (24)$$

$$m_H = 109.5 \pm 2.9 \ (5.2) \ GeV \text{ for } m_t = 171.4 \pm 3 \ (4)\sigma \ GeV, \sigma = 1.3 \ GeV. \quad (25)$$

However, the Higgs mass may be also obtained explicitly from the radiatively generated mass, as in Equ (3) with $z, k = 1$ and $y = 1 - \frac{1}{\pi}$,

$$m_H = m_t\sqrt{\frac{2}{\pi}\left(1 - \frac{1}{\pi}\right)} = m_t 0.6588 \cong m_t\frac{2}{3} = 115.4 \pm 0.9 \ GeV. \quad (26)$$

## 7. Higgs within the Type I transition

In this section, I hypothesize on physics that may be responsible for the k=1 branch Higgs masses, introduced in Section 3. I show that lighter Higgs mass solution is likely associated with the discontinuous Type I transition. In my terminology, effective theory abruptly changes the parameters and degrees of freedom across the low energy HMZC scale affiliated with the Type I transition.

I sketch class of models that may *exactly* remove tachyon solution at high energies and I introduce a more appropriate name for that transitional scale.

I analyze model structure, proposed by Popovic 2002 [33], where top quark is composite, composed of 3 fundamental fermions, and Higgs scalar is composite, composed of 2 fundamental fermions. I dub class of models, with $m_H \cong \frac{2}{3}m_t = 115.4 \pm 0.9 \ GeV$ as generic feature, the Composite Particles Model (CPM).

### 7.1. Composite Particles Model (CPM) and HMNZ^2 scale

#### 7.1.1. High energy limit

The low energy SM Higgs field may be created as a consequence of the electroweak symmetry breaking affiliated with the Type I transition. The "new physics", = a more fundamental physics, may conspire to promote effective scalar field with propagator free of leading divergences and with $\mu, \lambda = 0$ *at, just above* or maybe even *at all energies* above the low energy HMZC scale.

Calculation in 2D, Equ (3a) with $x = 1$, then leads to

$$g_{t*} = \sqrt{\frac{w}{12}(g_Y^2 + 3g_W^2)}\,, \tag{27}$$

i.e. the top Yukawa coupling at the HMZC scale and where $w = 1$ or $\frac{2}{3}$ is introduced to account for the correct counting of the gauge boson polarization degrees of freedom. In 2D one expect $w = 1$ for the longitudinal degree of freedom and in 4D one expect $w = \frac{2}{3}$ (1) for two transversal polarizations (all three polarizations). Traditionally, the gauge boson in 4D acquires the longitudinal polarization via the electroweak symmetry breaking; see for example [136].

Possibility that there are no tachyons *at all energies* above the HMZC scale $\sim \Lambda_{EWSB}$ but rather just massless particles whose exact zero mass is hence established by defining principle in Equ (27) or Equ (3a) with $x = 1$ ($y = 0$ or $\sqrt{\lambda} = 0$ or $m_H = 0$) seems worthwhile investigating. This would require a more appropriate name, the *Higgs Mass Non-Zero to Zero* (HMNZ^2) transitional scale instead of the HMZC scale. But, Higgs mass, then, needs to be stabilized *at all energies* larger than the HMNZ^2 scale!

The massless Higgs, within a non-zero VEV theory, one would suspect, should correspond to an ordinary, massless scalar at high energies in the early Universe, within a zero VEV theory. Therefore, if there are no other fundamental mass terms except Planck mass, $M_{Pl}$, one would imagine the Universe dynamics to be parameterized with a single mass parameter and set of couplings running naturally slow.

However, one may show that it is not possible to exactly remove tachyon solution with the unassisted SM at high energies. While quadratic and *all* logarithmic divergences in the scalar propagator cancel out at the HMZC scale thanks to Equ (27) and $\mu, \lambda = 0$, the gauge couplings and parameters defining Higgs potential are still running. Hence, nothing prohibits divergences to reappear at slightly larger energies.

However, let us still assume that Equ (27) may indeed describe the *decoupling limit* where scalar sector effectively decouples or disappears from the effective theory and tachyon solution is exactly removed. On those lines the Equ (27) needs to be matched with the low energy SM across the Type I transition.

Let us imagine that Equ (27) is satisfied for the low energy SM values of $g_Y, g_W$. One obtains

$$g_{t*} \cong 0.34\sqrt{w} \sim \frac{g_t}{3}\sqrt{w}\,. \tag{28}$$

However, in 4D, see Equ A5, the above results, i.e. Equ (27-28), translate to

$$w\frac{g_Y^2 + 3g_W^2}{4} - 2g_{t*}^2 = 0 \;\rightarrow\; g_{t*} = \sqrt{\frac{w}{8}(g_Y^2 + 3g_W^2)}\,, \tag{29a}$$

$$g_{t*} \cong 0.42\sqrt{w} \sim g_t \sqrt{\frac{w}{6}} = \frac{g_t}{3}\sqrt{\frac{3w}{2}}. \tag{29b}$$

As anticipated, Equ (27) and Equ (29a) with $g_{t*2D} = g_{t*4D}$ agree for $w = 1$ in 2D and $w = \frac{2}{3}$ in 4D, independent of the actual values of $g_Y$, $g_W$.

That is promising but there are at least three problems: (1) the SM $\lambda$ is not equal zero at low energies, (2) no SM fermion has that Yukawa coupling and (3) that Yukawa coupling is not strong enough to balance the QCD forces, i.e. based on discussion is Section 5, it appears that there is no suitable condensate that would break the electroweak symmetry.

One may compare the above situation with discussion in Section 4; dynamical generation of the appropriate masses for Z and W, e.g. based on the Pagels-Stokar relationship [114] or gap equations in the gauged NJL model [115,116], requires a large dynamically generated fermion mass; i.e. problem that motivated TC^2 [118-126] and Top Seesaw [127-129]. Unfortunately, the above situation suggests an even worse mismatch between the dynamically generated boson and fermion degrees of freedom!

But there is a cure.

### 7.1.2. *Composite Particles Model (CPM)*

First and second problems are indicative that this needs to be the Type I transition where parameters and degrees of freedom of the theory abruptly change across the HMZC scale. Second and third problems are suggestive that this might be interplay with at least three "fundamental" fermions forming a composite fermion, identified as top quark, with characteristics that provides condition for formation of top condensate responsible for the electroweak symmetry breaking, see section 5. This conclusion is anticipated by both 2D and 4D theories as result of Equ (28) and (29b).

If creation of composite or bound state is thought of to be a non-local phenomenon, as it should be, the Yukawa coupling should be thought of to be an additive quantity that adds to ~one as in the SM.

If there are 3 scalar fields within condensate of mass $2m_t$, the scalar mass, in the non-relativistic limit, is

$$m_H = \frac{2}{3}m_t = 115.4 \pm 0.9 \; GeV \tag{30}$$

which then appears, according to 2D considerations in Section 3, as the k=1 Higgs mass branch physics within the Type I transition at $\sim 10^{2.9} \cong 800 \; GeV$ [26].

In this picture, one expects the QCD assisted with a "new" physics to create (1) Higgs, "meson"-like particle consisting of two "fundamental" fermions, (2) top quark, "baryon"-like particle consisting of three "fundamental" fermions and (3) top condensates breaking the electroweak symmetry at the HMZC or HMNZ^2 scale. These lines of thoughts have been proposed in [33].

And Higgs mass in Equ (30) in the non-relativistic limit probably need to be corrected by only a relatively small amount; after all, the strong QCD interactions are not that strong above the $M_Z$!

I call this class of models, with Equ (30) as common feature, the Composite Particles Model (CPM).

As well known from the lattice arguments [94], the non-Abelian gauge field carries charge that cause its propagation to mimic 1 space dimensional flux providing confinement between static charges. This fact in connection with observed QCD confinement was original motivation for the string theories.

Finally, as emphasized by Nambu [111] the Higgs mass is determined as $m_H = 2m_t$ within the gauged Nambu-Jona-Lasinio mechanism when applied to the SM and implemented in the fermion loop approximation, see also [107]. In difference to that point of view, here, the top quark is expected to be composite too and the fundamental relationships defining CPM is expressed by Equ (30).

Therefore, in this picture, Equ (27) may describe the *decoupling limit* in a sense that there is no scalar field above the HMNZ^2 scale and therefore no tachyons, *exactly*, either; hence, similar to models with the strong interaction mediated dynamical EWSB [108, 109, 40]. Higgs mediated top-anti top interaction in Section 4 is identified here as a more complex dynamics associated with a "new" physics responsible for the CPM. Hence, top condensation may be a natural consequence of QCD and a "new" physics logarithmic running, i.e. resolving hierarchy problem and removing tachyons at once.

Alternative interpretation could be that there is still a composite scalar field but it is effectively decoupled from the electroweak sector of the theory.

On those lines, consider $\Phi\Phi \to GG$ scattering at energies above the HMNZ^2 scale where $G$ symbolizes electroweak gauge boson. By rewriting the longitudinal gauge bosons in terms of the Goldstone bosons and interpreting them as the fermion composites one may show that Lagrangian term for $\Phi\Phi \to GG$ scattering vanish if Equ (27) is satisfied. The right-handed fermion within one composite scalar field may couple with three left-handed fermions, either from another scalar field or from the two longitudinal gauge bosons, and the other right-handed fermion, belonging to other composite scalar field, may couple with other two left-handed fermions. That is factor of 6. The extra factor of 2 is due to the chiral symmetry which is unbroken at all energies above the HMNZ^2 scale. This distant version of the Goldstone Boson Equivalence Theorem applied to the CPM model will be addressed elsewhere.

The composite fermions have been addressed in the past [140], though in different context than here or [33]. More recently, the possibility that fermions may be composite is briefly discussed as one way of preventing the excessive FCNC in the "Littlest Higgs" model [141].

It was suggested, but without the large or medium scale UV completion, that the "Little Higgs" [141-144] may be composite if the sigma field is a condensate of strongly interacting fermions; hence the fermions may be composite with masses protected by approximate global symmetries [140, 141].

The models with composite scalars, with masses sensitive to the quadratic quantum corrections, have been proposed in the past [145-149]. Recent model building effort [141, 149] in the context of the $SU(5)/SO(5)$ breaking pattern [147, 148] and Higgs thought of as a light pseudo-Goldstone boson claimed that Higgs mass may be stabilized against radiative corrections. That is accomplished with approximate global symmetries, involving new heavy particles, that are imposed to soften the cutoff-dependence [141] and with TC^2 top-color interactions and conjecture that "we live in a region of the explicit chiral symmetry breaking interaction parameter space that lies between successive electroweak symmetry breaking phase transitions — at which $m_H$ and VEV must vanish" [149].

It would be interesting to investigate if these mechanisms may be compatible with CPM.

### 7.1.3. *Leading divergences at low energy*

Should there be cancelation of quadratic divergences at energies smaller than the HMZC scale? Or alternatively, what are degrees of freedom that might entangle that behavior at low energies?

Consider again the 2D theory with the Type I transition corresponding to $g_{t*} \sim \frac{g_t}{3} \to A g_{t*} \sim \frac{g_t}{3} A$ , $m_H = 0 \to m_H = \frac{2}{3} m_t$ and $\lambda = 0 \to \lambda = \frac{m_H^2}{v_{EW}^2} = \frac{4 m_t^2}{9 v_{EW}^2} = \frac{2 g_t^2}{9}$ and require cancelation above and below the HMZC, or HMNZ^2, scale. in 2D, The cancelation of leading divergences in 2D, Equ (A6), then leads to

$$\frac{g_Y^2 + 3 g_W^2}{4} = 3 g_t^2 \left(\frac{1}{3}\right)^2 \quad \text{and} \quad 3\lambda + \frac{g_Y^2 + 3 g_W^2}{4} = \frac{6 g_t^2}{9} + 3 g_t^2 \left(\frac{1}{3}\right)^2 = 3 g_t^2 \left(\frac{1}{3}\right)^2 A^2 \quad (31)$$

where Equ (30) was assumed and factor A was introduced as a free parameter to be determined. I find

$$A^2 = 3 \qquad (32)$$

which is the k=1 solution with $y = \mathbf{0.66}$ introduced in Section 3. Therefore, the above Type I transition is simply $x = 1 \to x = \mathbf{0.33}$ transition, in the notation of Section 3, and taking place at HMZC, or HMNZ^2, scale $\sim \Lambda_{EW} \sim 10^{2.9} \cong 800 \; GeV$ [26].

After inspection, one finds the top quark mass as in [33],

$$g_t = \sqrt{3 \frac{g_Y^2 + 3 g_W^2}{4}} \cong 1.025 \; \to \; m_t = \sqrt{\frac{3}{2}(M_Z^2 + 2 M_W^2)} = 178.51 \; GeV. \qquad (33)$$

What about the 4D cancelations?

Well, left hand expression in (31) is identical in 2D and 4D for $w = \frac{2}{3}$. And 4D analogy of the right hand expression in Equ (31) maybe *do not need* to be satisfied!

If Higgs mechanism takes place dominantly in 2D then VEV do not need to be non-zero everywhere in space-time which clearly removes both the hierarchy and vacuum energy problems. Therefore the EWSB may be determined by 2D propagator physics and 4D couplings. For low energy HMZC or HMNZ^2 scale there should be no substantial fine tuning. I revisit this problem in Section 7.4, where I show that it is possible, after all, to cancel the leading divergences and conserve theory structure across the space-times with different dimensions, both below and just above the HMZC, or the HMNZ^2 scale.

Next, I investigate the "new" physics imprints that could reproduce the above CPM structure by sketching several models that deal with external particle degrees of freedom within the 2D and 4D space-times as well as with degrees of freedom (e.g. color, flavor) in the internal space.

## 7.2. $3^{-1/2}$ model with flavor

Consider that the right handed *up* quarks create condensates in a generation universal manner, i.e. each of the right handed *up* quarks couples with each of the left handed *up* quarks. Furthermore, each of the nine condensates is assigned scalar field $\Phi_{ij}$ where $i,j = 1,2,3$ and, finally, each scalar acquires identical vacuum expectation value $v_{EW}/3$. However, there exists only one non-zero fermion mass eigenstate which corresponds to the identically populated superposition of the left handed and the right handed "original" states. That massive fermion is identified as the top quark. The top condensate will have non-zero VEV that is three times larger than the "original" condensate value,

$$\langle \bar{\Psi}_t \Psi_t \rangle = \langle \frac{1}{\sqrt{3}}(\bar{\Psi}_1 + \bar{\Psi}_2 + \bar{\Psi}_3) \frac{1}{\sqrt{3}}(\Psi_1 + \Psi_2 + \Psi_3) \rangle = 3 \langle \bar{\Psi}_i \Psi_j \rangle \tag{34}$$

where $i,j = 1,2,3$. And the other two fermion mass eigenvalues are zero.

The top quark however couples with *three* physical scalar fields below the HMZC scale, each of which is superposition of the scalar fields corresponding to the condensates with common left-handed partner

$$\Phi_i = \sqrt{\Phi_{i1}^2 + \Phi_{i2}^2 + \Phi_{i3}^2} \leftrightarrow \bar{\Psi}_{iL}\Psi_{jR} + \bar{\Psi}_{jR}\Psi_{iL} \text{ where } j = 1,2,3. \tag{35}$$

Superposition which mixes the left-handed partners is meaningless due to $SU(2)$ rotations that mix the final fermion mass eigenstates. Each of the three fields $\Phi_i$ where $= 1,2,3$, acquires $v_{EW}/\sqrt{3}$, whereas the physical top quark couples to each of them with coupling $3^{-1/2}$. Interested reader may note that this physics may be described with the effective Lagrangian density in Equ (14) addressed in Section 6.3.

If one now assumes the CPM non-relativistic limit $m_H = \frac{2}{3} m_t$, which I base on discussion in Section 7.1, and consider leading divergences in 2D with $w = 1$ and Type I transition with $\frac{g_t}{3} \to \frac{g_t}{\sqrt{3}}$, $\lambda = 0 \to \frac{2}{9}$ i.e. $m_H = 0 \to \frac{2}{3} m_t \in (115.4, 119.0) GeV$ while crossing the HMZC, or HMNZ^2, scale from higher to lower energies, one finds the conditions for cancelation for scalar fields $\Phi_{ij}$ and $\Phi_i$ respectively to be

$$\frac{g_Y^2 + 3g_W^2}{4} = 3g_t^2 \left(\frac{1}{3}\right)^2 \text{ and } 3\lambda + \frac{g_Y^2 + 3g_W^2}{4} = \frac{6g_t^2}{9} + \frac{g_Y^2 + 3g_W^2}{4} = 3g_t^2 \left(\frac{1}{\sqrt{3}}\right)^2 \tag{36}$$

i.e. two numerically identical expressions corresponding to two completely different physical interpretations. Hence, one obtains the prediction for the top quark Yukawa coupling and mass as [33]

$$g_t = \sqrt{3 \frac{g_Y^2 + 3g_W^2}{4}} \cong 1.025 \to m_t = \sqrt{\frac{3}{2}(M_Z^2 + 2M_W^2)} = 178.51 \, GeV. \tag{37}$$

In 4D with $= \frac{2}{3}$, the left hand side of Equ (36) takes the form

$$w(M_Z^2 + 2M_W^2) = 4m_t^2 \cdot \left(\frac{1}{3}\right)^2 \to m_t = 178.51 \, GeV. \tag{38}$$

Hence, one again obtains Equ (37). The right hand side of Equ (36) in 4D with $w = \frac{2}{3}(1)$ takes the form

$$m_H^2 + w(M_Z^2 + 2M_W^2) = 4m_t^2 \left(\frac{1}{\sqrt{3}}\right)^2 \rightarrow m_H = 160.6 \; GeV \; (136.8 \; GeV) \qquad (39)$$

for $m_t = 173.1 \; GeV$. This translates to $m_H = 168.3 \; GeV \; (145.8 \; GeV)$ for $m_t = 178.1 \; GeV$, as suggested by Equ (37); result that is clearly inconsistent with the $m_H = \frac{2}{3}m_t$ premise. But as discussed above, Equ (39) *do not need* to be satisfied at energies smaller than the low energy HMZC, or the HMNZ^2 scale.

The $w = 1$ solution introduced above appears as the k=2 branch Higgs mass in the close vicinity to two special Planck mass affiliated solutions, discussed in Section 6.1. The *approximate* $\mu, \lambda = 0$ solution [33] is obtained for the top quark mass world average, $m_t = 173.1 \; GeV$, while Higgs within the Planck mass version of the Coleman-Weinberg conjecture [34] is obtained for the non-relativistic limit prediction $m_t = 178.1 \; GeV$, where $m_t$ is obtained from the Z and W gauge boson masses alone.

The $w = \frac{2}{3}$ mass, Equ (39), extends beyond the $147 \; GeV$ limit, requiring a single HMZC scale at energies smaller than the Planck mass, addressed in Section 2, but it is below the $171 \; GeV$ perturbativity limit. However, the revisited global electroweak fit [48] excludes this solution. Similarly, the recent Tevatron analysis [49-51] excluded the Higgs within the $162 \; GeV < m_H < 166 \; GeV$ range at 95% C.L.; whereas the entire $160 \; GeV < m_H < 170 \; GeV$ range appears unlikely. Hence, the $w = \frac{2}{3}$ solution is unlikely.

It would be rather out of the ordinary situation if the late LEP suspicious signals [44] corresponds to important 2D Higgs dynamics centered at $m_{H2D} \cong 117 \; GeV$, whereas the LHC discovers 4D Higgs dynamics centered at $m_{H4D} \cong 141 \; GeV$ for $w = 1$ case (all three polarizations), as described above.

The combined CDF and DØ analysis [49] observed the $1\sigma$ excess for $132 \; GeV < m_H < 143 \; GeV$ Higgs.

I revisit the potential interplay between the 2D and 4D Higgs dynamics in Section 7.4.

### 7.3. $3^{-1/2}$ model with color

Consider model with 9 scalars that mix fermions with different colors, but not flavor, at energies slightly larger than the HMZC, or the HMNZ^2 scale. The fermion top Yukawa coupling equals $\frac{g_t}{3}$. However, at energies slightly smaller than the HMZC, or the HMNZ^2 scale each top color couples with color specific Higgs scalar particle and there are three of those. Hence, each scalar field acquires $v_{EW}/\sqrt{3}$ and again the physical top quark couples to each of them with coupling equal to $g_t 3^{-1/2}$. This model structure, which is numerically identical to those presented in Sections 6.3 and 7.2, was proposed in [33].

If one again assumes non-relativistic CPM limit with $m_H = \frac{2}{3}m_t$, which I again base on discussion in Section 7.1, and consider leading divergences, one should discover, as in Section 7.2, the Type I transition where $\frac{g_t}{3} \rightarrow \frac{g_t}{\sqrt{3}}$, $\lambda = 0 \rightarrow \frac{2}{9}$ i.e. $m_H = 0 \rightarrow \frac{2}{3}m_t \in (115.4, 119.0)GeV$. This Type I transition is consistent with result of analysis that led to Equ (31-32).

I now consider quadratic divergences in 2D, with $w = 1$, above and below the HMZC or HMNZ^2 scale. Condition for cancelation above the HMZC, or the HMNZ^2 scale for 9 scalar fields is

$$\frac{g_Y^2+3g_W^2}{4} = 3g_t^2\left(\frac{1}{3}\right)^2 .\tag{40}$$

In other words, there is a single color combination in the fermion loop and fermion coupling is $g_t/3$.

Condition for cancellation beneath the HMZC or HMNZ^2 scale for 3 color specific scalar fields is

$$3\lambda + \frac{g_Y^2+3g_W^2}{4} = \frac{6g_t^2}{9} + \frac{g_Y^2+3g_W^2}{4} = 3g_t^2\left(\frac{1}{\sqrt{3}}\right)^2 \tag{41}$$

where there is again a single color specific combination in the fermion loop but coupling is now $g_t 3^{-1/2}$.

After inspection one should discover that Equ (40) and Equ (41) are numerically identical.

Hence, one again obtains prediction of the top Yukawa coupling [33] that is identical to Equ (37) in Section 7.2., result that is 4 "$world$" $\sigma$ (less than 3%) away from the world average top quark mass [60].

Reproducing the above analysis in 4D would suggest the same result as in Equ (38-39), see Section 7.2.

### 7.4. *2D and 4D models with movers*

Imagine for a moment that the 2D "fundamental" fermions are simply the x, y, and z movers. The top quark appears as composite of all three species. Therefore, the 2D "fundamental" fermions couple with strength as in Equ (28). Here I assume that phase space for top loop at energies smaller than the HMZC has an additional factor $b = 3$ or $\pi$ to account for dynamics in the entangled, orthogonal 2D space when described in terms of a "new" physics = effective theory at energies larger than the HMZC scale.

Therefore a "new" physics, defining the "original" coupling $g_{t*}$, transforms, at energies slightly smaller than the HMZC, or the HMNZ^2 scale to the minimal SM with broken electroweak symmetry, as parameterized with a set of the effective parameters, and reproducing the CPM structure,

$$g_{t*} \to g_{t*}\sqrt{b}, m_H = 0 \to m_H = \frac{2}{3}m_t \text{ and } \lambda = 0 \to \lambda = \frac{m_H^2}{v_{EW}^2} = \frac{2g_t^2}{9} .\tag{42}$$

Again, consider the leading divergences in 2D and require that gauge boson loops are canceled at energies slightly larger than the HMZC, or the HMNZ^2 scale, whereas both the gauge and Higgs boson loops are cancelled at energies smaller than the HMZC, or the HMNZ^2 scale, i.e. according to Equ (A6),

$$3\lambda + \frac{g_Y^2+3g_W^2}{4} = \frac{6g_t^2}{9} + 3g_{t*}^2 = 3g_{t*}^2 b \tag{43}$$

$$\to \quad \frac{g_t}{g_{t*}} = 3\sqrt{\frac{1}{2}(b-1)} = 3.00\ (3.10) \text{ for } b = 3\ (\pi). \tag{44}$$

Now, I will match the 4D model with the above 2D model structure. I use scaling obtained in Section 6.3., i.e. $\frac{\lambda_{4D}}{\lambda_{2D}} = \left[\frac{2}{3} - 3\left(x_1 - \frac{y_1}{\pi}\right)\right]/\left(\frac{y_1}{\pi}\right) \cong 1.47$ . Therefore, Equ (43-44) translate in 4D to

$$1.47\lambda + \frac{g_Y^2+3g_W^2}{4} = 1.47\frac{2g_t^2}{9} + \frac{2g_{t*}^2}{w} = 2g_{t*}^2 b \tag{45}$$

$$\rightarrow \quad \frac{g_t}{g_{t*}} = 3\sqrt{\frac{b-w^{-1}}{1.47}} = 3.03 \ (3.17) \text{ for } w = \frac{2}{3} \text{ and } b = 3 \ (\pi). \tag{46}$$

And only the $w = \frac{2}{3}$ solution is consistent with the premise with 3 "fundamental" fermions whose Yukawa couplings add linearly. This is as expected, see Equ (29). It is traditionally anticipated that the gauge bosons in 4D have two transversal polarizations in the unbroken electroweak phase and subsequently gain one additional, longitudinal, polarization in the broken electroweak phase.

If there are $\pi$ "fundamental" fermions then Equ (44) and (46) suggest consistent structure for $b = \pi$.

Therefore, discovered theory is within the Type I transition with abrupt change of parameters defined by Equ (42) with cancelation of quadratic divergences in both 2D and 4D, with correct counting of the gauge bosons polarizations and almost consistent ratio equal 3 between the physical and "bare" top Yukawa couplings: 3.03 and 3.00 in 4D and 2D respectively.

Hence, this model structure may exactly remove tachyons both in the fundamental 2D and 4D theories!

I was able to retain the 4D cancellation thanks to the appropriately understood scaling between the 2D and 4D theories. Clearly, this may be even applied to models with different physical interpretation.

The Higgs mass and top quark mass in 2D model equal

$$m_H = \frac{2}{3} m_t \ (115.4, 119.0) \, GeV, \, m_t \in \left(173.1, \sqrt{\frac{3}{2}(M_Z^2 + 2M_W^2)} = 178.5 \, GeV\right) \tag{47}$$

corresponding to the phase transition at $\sim 10^{2.9} \cong 800 \, GeV$ [26,33]. If predicted top quark mass is scaled back to the world average mass and predicted Higgs mass is scaled by the same factor, one obtains

$$m_H \rightarrow \frac{173.1}{178.5} 115.4 \, GeV = 111.9 \, GeV. \tag{48}$$

Similarly, the Higgs mass and top quark mass in 4D model equal

$$m_H = \sqrt{1.47 \frac{2g_t^2}{9}} \, v_{EW} = \sqrt{1.47 \frac{2}{9} 3.03^2 g_{t*}^2} \, v_{EW} = \sqrt{1.47 \frac{2}{9} 3.03^2 \frac{g_Y^2 + 3g_W^2}{12}} \, v_{EW},$$

$$\rightarrow \quad m_H = \sqrt{1.47 \frac{2}{9} 3.03^2 \frac{M_Z^2 + 2M_W^2}{3}} = 145.8 \, GeV,$$

$$m_t = \sqrt{3.03^2 \frac{M_Z^2 + 2M_W^2}{6}} = 180.3 \, GeV. \tag{49}$$

with the electroweak phase transition scale roughly in the range $1 - 1.15 \, TeV$.

If predicted top quark mass is scaled back to the world average top quark mass and predicted Higgs mass is scaled by the same amount, then one obtains

$$m_H \rightarrow \frac{173.1}{180.3} 145.8 \, GeV = 140.0 \, GeV \tag{50}$$

in the close vicinity of the $m_H = 138.1\ GeV$ solution, see [33], discussed in Section 6.1.

It would be rather out of the ordinary situation if the late LEP suspicious signals [44] corresponds to important 2D Higgs dynamics centered at $m_{H2D} \cong 117\ GeV$, whereas the LHC discovers 4D Higgs dynamics centered at $m_{H4D} \cong 140\ GeV$, for $w = \frac{2}{3}$, i.e. for only two, transversal, polarizations of the 4D massless gauge bosons in the high energy regime as described above.

## 8. Conclusion and summary

With LHC collecting the high energy data it is expected that our understanding of Nature will dramatically advance in the near future. Paradigm shifts in physics always generated many important new technologies with a vast range of practical applications. Hence, there is a good chance that society will again benefit greatly from this largest physics research endeavor ever undertaken by mankind.

Why do we expect a huge paradigm shift in the next year or two? Well, there is a single particle anticipated by the current SM dogma [1-21] awaiting to be discovered. And it seems certain that the SM Higgs should be within the reach of the LHC. But it is also generally accepted that current dogma is incomplete and incorrect; the reason being hierarchy [40] and vacuum energy problems [41-43].

Will the LHC address physics outside of the current dogma? As I show in this paper, it must. Even in the next to the "worst case" scenario, where Higgs is discovered and there is nothing else to surprise us at small energies, there is still an energy scale, HMZC $\sim \Lambda_{EWSB}$ [26, 33], see also Section 2, within the LHC reach where current dogma itself suggests that effective Higgs particle should transition from standard particle, positive mass squared, to tachyon, negative mass squared, degree of freedom. Never in history have particle physicists dealt with anything similar. In April this year, the LHC reached the CM energies ($\sim 7\ TeV$) that embraces the entire HMZC range. Therefore, yes, the LHC is just beginning to make its mark in history by stepping distinctively outside of the known physics territories.

By preparing system the average CM collision energies can be brought to the HMZC energy $\sim \Lambda_{EWSB}$; corresponding to condition for the phase transition which probably happened in the very early Universe [70], though in opposite direction and without the actual change of the ground state. As discussed, by going back in physical time the ground state of today's Universe with non-zero VEV transitions to one with zero VEV. And in the zero VEV Universe the tachyon Higgs is just an ordinary non-tachyon scalar particle. Therefore, we might learn a lot about the actual EW phase transition by studying physics at and around the HMZC scale. While tachyon theories are often addressed in the context of string theory and cosmology, there is an *alarming lack* of literature and ongoing research effort among the rest of particle physics community. I hope that this paper will motivate more focused research effort in that direction.

In this paper I carefully map the physical Higgs mass with the low energy HMZC scale $\sim \Lambda_{EWSB}$ based on my earlier work back in 2001 [26] and 2002 [33]. As shown here, the HMZC scale exist for the Higgs mass lighter than approximately 200 GeV; range that is also strongly favored by the electroweak precision data and direct Higgs searches [46]; see Section 2.2. The general HMZC scale range is $10^{2.9} - 10^{3.7}\ GeV$ ($800\ GeV - 5\ TeV$). For the Higgs masses within the $114.4 < m_H < 182\ GeV$ range [46], preferred by global fit and direct searches, the HMZC scale range is $800\ GeV - 1.8\ TeV$, see Fig 2.

If the SM is expected to be valid at all energies smaller than the Planck mass, i.e. within the SM "desert" or "long lived" scenario, then renormalization group flow implies that Higgs must be heavier than $137.0 \pm 1.8\ GeV$, based on the vacuum stability limit [26, 52-59, 46]; otherwise, unacceptable deeper minimum of the effective potential occurs. Similarly, the SM renormalization group flow implies that Higgs mass must be lighter than $171 \pm 2\ GeV$, based on the perturbativity limit [26, 60, 61, 46]; otherwise, scalar self interactions diverge, i.e. strongly coupled Higgs sector cannot be described with the perturbation theory. Finally, because there are generally two HMZC scales per physical Higgs mass, the condition that there is a single HMZC scale [26, 33] at energies smaller than the Planck mass puts an upper limit on the Higgs mass equal to $146.5 \pm 2\ GeV$. Therefore, if the SM is a valid description of Nature at all energies below the Planck scale, where it has the effective structure of an unbroken electroweak symmetry, then the stability curve and condition that there is a single HMZC scale at all energies smaller than the Planck scale limits the SM Higgs mass to a very tight range of roughly $142 \pm 6\ GeV$ with a corresponding electroweak phase transition scale roughly in the range $1 - 1.15\ TeV$.

However, traditionally, there is no strong enough reason to expect the SM to be a valid theory at all energies smaller than the Planck scale. Just to the contrary, hierarchy problem [40] affiliated with the presence of leading quantum corrections associated with quadratic divergences in the scalar propagator suggests that theory cannot be valid across vast energy scales unless there is a defining principle that provides explanation which goes beyond the current dogma. Otherwise, the theory quickly, i.e. already after couple of magnitudes in energy, becomes unnaturally finely tuned.

In Section 2.5, I present an analysis which compare the HZMC scale with stop mass to show that the MSSM [27-31] is less unnatural than the SM at low energies for $m_H \leq 120.9 \pm 0.9\ GeV$.

In Section 3, I investigate the leading SM quantum corrections to the scalar propagator in 2D theory. I show that one could simultaneously satisfy (1) complete radiative generation of the Higgs mass through top loop and (2) complete cancelation of the remaining leading quantum corrections to scalar propagator. There is unique solution for the Higgs mass. This solutions is parameterized with k=1 or 2 and the corresponding SM solutions in the zeroth order are $113.0 \pm 1.0\ GeV$ and $143.4 \pm 1.3\ GeV$. I addressed three types of uncertainties that could probably push these results by couple $GeV$'s.

It is worth noting that the k=1 branch almost embraces the late LEP suspicious signals [23-24, 44-46] in the vicinity of $115\ GeV/c^2$ whereas the k=2 branch almost embraces the "long lived" SM solutions, I discuss in Section 6.1, in the context of the SM renormalization flow all the way up to the Planck mass. Interestingly, the combined CDF and DØ analysis [49] observed $1\sigma$ excess for $132\ GeV < m_H < 143\ GeV$.

Both hierarchy and vacuum energy problems may be eradicated If EW symmetry breaking takes place in 2D which "embeds" the physical propagation in 4D. The leading divergences in 2D are only logarithmically divergent and therefore the large hierarchy may be easily attained. Moreover, the Higgs scalar field doesn't need to be non-zero constant in the entire 4D space hence leading to reasonable 4D space-time curvature. As known by the lattice arguments [94], the non-Abelian gauge field carries charge that cause its propagation to mimic 1 space dimensional flux providing confinement between static charges. Hence, I hypothesize that 4D description may be only an effective 2D description as this can solve hierarchy and vacuum energy problems. This could be to the extent that: (1) 4D electroweak symmetry breaking is governed by 2D electroweak symmetry breaking and 4D couplings or (2) that 4D theory is effective theory completely described by 2D theory, where dimensionality of space-time enters less as a premise and more as a consequence of the fundamental 2D theory.

Finally, independent of the space-time dimensionality, I suggest that the EW symmetry breaking may be both dynamical and *local* in sense that effective Higgs scalar field might be zero in almost entire space except in the close vicinity of physically propagating massive particles.

The Technicolor theories [108, 109, 40] were introduced to address the hierarchy problem [40]; main idea was that there may be additional forces beyond the SM that could provide glue for fermions to bond and break the EW symmetry. The entire point was that gauge couplings associated with these forces would run logarithmically and therefore provide natural explanation for large hierarchy.

In Section 5, I show that top condensate formation may be consistent with interplay between the QCD gluon and Higgs scalar mediated top anti-top interactions. Vice versa, starting with this as a premise I predict the fine structure constant of the strong QCD interactions up to precision better than 2% in the leading order calculation and explain how to reach even better agreement. Interestingly, the predicted mean value is only 0.25% away from the world average value at $\sqrt{s} = M_Z$. Therefore it seems that dynamical top condensation may indeed be a viable option with either fundamental or composite Higgs.

As previously noted [33] there is one solution of the SM renormalization flow in the vicinity of $138.1\ GeV$ Higgs that is a very distinctive one. Both the effective Higgs mass and the quartic scalar coupling conspire to be almost zero at the same high energy scale. This high energy scale happens to be in the close vicinity to the Planck mass scale. Furthermore, for both $\mu$ and $\lambda$ the SM renormalization flow is rather logarithmic and already based on visual inspection, see Fig 7, it appears that these quantities are directly related. Another distinctive solution, centered at $146.5\ GeV$, corresponds to the Planck scale version of the Coleman-Weinberg (CW) conjecture [52].

In Section 6, I address the composite Higgs built out of top quark degrees of freedom at very high energies. I show that the top-built composite Higgs has an effective mass squared proportional to the square root of quartic coupling also equal to the top Yukawa coupling and, therefore, naturally seems to imply a "long lived" solution. This observation is bound to energies larger than the low energy HMZC scale. The fact that Higgs mass squared is positive at low energies likely means that quantum corrections due to additional dynamics (most likely affiliated with top condensation physics) overcome the leading order prediction of the negative Higgs mass squared within the top-quark built composite Higgs model. While explaining hierarchy, this model in 4D doesn't explain the vacuum energy problem and therefore the hypothesis about 2D fundamental theory still holds. In a sense, it should not be surprising that this solution corresponds to the k=2 branch I discovered within the 2D theory considerations.

In Sections 6 and 7, I address the possible physics realizations beyond the standard dogma [1-21]. I introduce several conservative and a few radical models that deal with both external and internal degrees of freedom. Section 6 is mostly concerned with the Type II transition where, in my terminology, the effective theory does not abruptly change the parameters and degrees of freedom across the low energy HMZC scale. Whereas, Section 7 is mostly concerned with the Type I transition where the effective theory abruptly changes the parameters and degrees of freedom across the HMZC scale.

In Section 7, a new class of models is introduced within the Type I transitions that are neither Supersymmetric [27-31], Technicolor [108, 109, 40], Topcolor [117], TC^2 [90-98], Top Seesaw [127-129] or Little Higgs [141-144] alike. All divergences (logarithmic terms are also zero) between fermions and electroweak gauge bosons loops in the Higgs propagator at energies in the vicinity of the HMZC scale are exactly cancelled therefore *exactly* removing the tachyon solution. Removing tachyon *at all energies larger than* the HMNZ^2 scale however likely requires dynamical and maybe *local* symmetry breaking and composite Higgs beyond the SM, see Section 7.1. Standard leading order cancelation is obtained at energies smaller than the HMZC or HMNZ^2 scale. Different levels of physical granularity below and above $\Lambda_{EW}$ are described with the Type I transition $x = 1 \rightarrow x = 0.33$ in the notation of Section 3, or $g_{t*} \sim \frac{g_t}{3} \rightarrow \sqrt{3} g_{t*} \sim \frac{g_t}{\sqrt{3}}$, $m_H = 0 \rightarrow m_H = \frac{2}{3} m_t$ and $\lambda = 0 \rightarrow \lambda = \frac{m_H^2}{v_{EW}^2} = \frac{4 m_t^2}{9 v_{EW}^2} = \frac{2 g_t^2}{9}$ in the SM notation.

| Model | $m_t$ | $m_H$ | $m_H (GeV)$ |
|---|---|---|---|
| k=2 branch | $173.1 \pm 1.3\ GeV$ | $\sqrt{\frac{6 m_t^2 (z) - M_Z^2 - 2 M_W^2}{3 \left(1 + \frac{\pi}{2}\right)}}, z = 1 \left(\frac{2 m_t^2}{V_{EW}^2}\right)$ **(2D)** | $143.4 \pm 1.3$ $(142.5 \pm 2.4)$ |
| Long lived @ Planck | $173.1 \pm 1.3\ GeV$ | $\cong 138.5\ GeV$ from $\mu$ & $\lambda$ **(4D)** $\cong 146.0\ GeV$ from $\mu$ **(4D)** | $138.1 \pm 1.8$ $146.5 \pm 2.0$ |
| $3^{-1/2}$ (k=2) | $173.1 \pm 1.3\ GeV$ | $\sqrt{\frac{4}{3} m_t^2 - M_Z^2 - 2 M_W^2}$ **(4D)** | $136.8 \pm 2.2$ |
| k=1 branch | $173.1 \pm 1.3\ GeV$ | $\sqrt{\frac{6 m_t^2 (z) - M_Z^2 - 2 M_W^2}{3(1 + \pi)}}, z = 1 \left(\frac{2 m_t^2}{V_{EW}^2}\right)$ **(2D)** and scaled to 4D with $1.4676 = \left[\frac{2}{3} - 3 \left(x_1 - \frac{y_1}{\pi}\right)\right] / \left(\frac{y_1}{\pi}\right)$ | $113.0 \pm 1.0$ $(112.3 \pm 1.9)$ $136.9 \pm 1.3$ $(136.0 \pm 2.3)$ |
| Movers 2D 4D | $\sqrt{3 \frac{3}{2\sqrt{2}} \frac{M_Z^2 + 2 M_W^2}{6}}$ **(4D)** $\sqrt{\frac{3}{2}(M_Z^2 + 2 M_W^2)}$ **(2D)** | $\sqrt{\frac{1}{3} \frac{M_Z^2 + 2 M_W^2}{3}}$ **(4D)** $\frac{2}{3} m_t$ **(2D)** | 145.3 (140.0) 115.4 (111.9) 119.0 (114.6) |
| $3^{-1/2}$ (k=1) Flavor Color | $\sqrt{\frac{3}{2}(M_Z^2 + 2 M_W^2)}$ **(2D & 4D)** | $\frac{2}{3} m_t$ | $115.4 \pm 0.9$ $119.0$ $115.4 \pm 0.9$ $119.0$ |
| $\pi^{-1/2}$ | $173.1 \pm 1.3\ GeV$ | $\sqrt{\frac{1}{3} \left(\frac{6}{\pi} m_t^2 - M_Z^2 - 2 M_W^2\right)}$ **(2D)** & $m_t \sqrt{\frac{2}{\pi}\left(1 - \frac{1}{\pi}\right)} \cong \frac{2}{3} m_t$ **(2D)** | $109.5 \pm 1.4$ $114.1 \pm 0.9$ |

**Table 1 The Higgs mass predictions. If Higgs mass below 114 $GeV$ is excluded the obtained values are roughly centered at $116.5\ GeV$, typically associated with the Type I transition with a single exception, and at $140.5\ GeV$, typically associated with the Type II transition with a single exception.**

A new class of models, dubbed Composite Particles Model (CPM) has 3 fundamental fermions creating a composite particle that is identified as top quark, whereas 2 fundamental fermions create Higgs. The non-relativistic limit in the "weak" QCD regime suggests $m_H = \frac{2}{3} m_t = 115.4 \pm 0.9 \, GeV$.

The Higgs compositeness within CPM appears very different from the Higgs compositeness in the context of $SU(5)/SO(5)$ breaking pattern [147, 148]. However, it would be interesting to investigate if there is a way to connect these two composite frameworks.

The summary of the Higgs mass predictions is presented in Table 1. The obtained values are roughly centered at $116.5 \, GeV$, typically associated with the Type I transition with one exception, and at $140.5 \, GeV$, typically associated with the Type II transition with a single exception.

The theme, which clearly stands out for the Type I transition, is too dominant top loop; the expected "balance" between electroweak gauge bosons and top loops in the unbroken phase exists for $\cong 3$ times smaller physical top Yukawa coupling. This observation persist in both 2D and 4D theories. This creates an even stronger mismatch between dynamically generated boson and fermion degrees of freedom than what can be observed by Pagels-Stokar [114] or gap equations in the gauged NJL model [115, 116] formalism, the problem that motivated TC^2 [118-126] and Top Seesaw [127-129] models.

One could resolve this problem, and possibly remove tachyons, by introducing additional phase space in the scalar sector hence, lowering the top Yukawa coupling.

Or, one could consider Composite Particles Model (CPM), [33], in which 3 fundamental fermions create a composite particle that is identified as top quark whereas 2 fundamental fermions create Higgs. The non-relativistic limit in the relatively "weak" QCD regime suggests $m_H = \frac{2}{3} m_t = 115.4 \pm 0.9 \, GeV$. Furthermore, in the zeroth order one could relate the top quark mass to the values of Z and W boson masses alone and obtain prediction of the top quark mass as in Popovic 2002 [33]. This prediction is less than 3% larger than the current world average top quark mass [69].

It should not be surprising if both Higgs and top quark masses in the next to leading order vary by up to several percents. Furthermore, adding additional structure that could, for example, explain the masses of other particles, could probably account for a few additional %s.

Therefore, a more detailed investigation of this class of models is necessary. It rarely happens that over-constrained system provide such good matches with observations.

As well known from the lattice arguments, e.g. see [94], the non-Abelian gauge field carries charge that causes its propagation to mimic the 1-space dimensional flux providing confinement between static charges. While QCD confinement motivated early phenomenological string theories, the above argument relaxed a need for the low energy phenomenology involving fundamental strings. Hence, in the similar spirit, the 2D considerations presented in this paper might be just an effective description, a consequence of complex dynamics of the non-Abelian gauge fields in the regular 4D space-time.

As emphasized by Nambu [111] the Higgs mass is determined as $m_H = 2m_t$ within the gauged Nambu-Jona-Lasinio mechanism when applied to the SM and implemented in the fermion loop approximation, see also [107]. In difference to that point of view, here, the top quark is expected to be composite particle and the fundamental relationships defining CPM is expressed by $m_H = \frac{2}{3} m_t$.

In Section 6.3, I obtain an important scaling between the 2D and 4D theories. Thanks to that scaling, the theory structure was conserved across space-times with different dimensions as well across two different regimes below and above the HMZC scale, see section 7.4.

Finally, the general concept of Higgs tachyon solution above the HMZC scale requires much better understanding. According to Wigner [138], the space-like negative mass squared particles have non-compact little groups so their spin is not described by rotation group $SU(2)$ and in difference to massless particles their "spin" may be continuous parameter. Maybe that fact could be related to the observed mismatch or "imbalance" between electroweak gauge bosons and top loops across the HMZC scale.

In this paper, I point out to (1) the importance of the effective 2D SM description in relation to the current problems that particle physics faces, (2) the possibility that QCD gluon and Higgs sectors may be closely related to the top condensation, and (3) the two regions of the theoretically preferred Higgs mass with accompanying models, see also [33].

I present class of composite models with dynamical and likely *local* symmetry breaking, which I dub the Composite Particles Model (CPM), which may exactly remove the tachyon solution at the high energies, i.e. energies larger than the HMZC or HMNZ^2 scale. Finally, I map the physical Higgs mass with HMZC scale $\sim \Lambda_{EWSB}$ based on my earlier work in 2001 [26] and 2002 [33].

As I "proved" in this paper the LHC experiment already stepped outside the known physics territories and one should definitively expect answers that go beyond the SM dogma [1-21] in the near future. The HMZC scale is better motivation for new physics than the perturbative unitarity consideration [35] which only motivates existence of standard Higgs sector at low energies but it does not necessarily motivate *new physics beyond the regular SM* Higgs at low energies.

The assumption here is that the LHC experiment will not change the meta-stable ground state; as generally accepted, changing the meta-stable ground state, if something like that exist at all, may not be a particularly wise thing to do. The catastrophic false vacuum scenario has been addressed by Coleman [80] and Callan and Coleman [81]. Many authors also addressed the meta-stable vacuum [82-91].

In similar context, and as precautionary measure, probably the most responsible thing to do would be to have a list with various worst case theoretical scenarios, not excluded by the ultra-high energy cosmic ray arguments, see [92] for review and reference therein, a list of the corresponding experimental signatures and finally an automatic real time detection / shut down system.

This author is not aware of any other more appropriate approach.

# Appendix

Here, I overview some of details of analysis [26, 33] regarding vacuum stability [52-59, 46], perturbativity [60, 61, 46] and the scale $\Lambda_{HMZC} \sim \Lambda_{EWSB}$ affiliated with the physical electroweak phase transition. More details are provided in the original publication [26] and references therein.

In parallel, two independent techniques were utilized: the $\overline{MS}$ scheme [62], applied to the effective potential [34] analysis [63-65], and the Euclidean hard cut-off scheme, applied to the generalized original Veltman's approach [66], confirmed by Osland and Wu [67] and added with logarithmic divergences by Ma [68].

Using a notation adopted here, the tree level potential of the neutral component of the Higgs scalar doublet is

$$V(\Phi) = -\frac{m_H^2}{4}\Phi^2 + \frac{\lambda}{8}\Phi^4 \tag{A1}$$

where $m_H^2 = V^{(2)}|_{\langle\Phi\rangle}$ is the tree level Higgs mass squared and $\langle\Phi\rangle = v_{EW} = 246.2\ GeV$. Hence, the running effective potential is defined as

$$V(\Phi_R) = -\frac{m_H^2(\Lambda \sim \Phi_R)}{4}\Phi_R^2 + \frac{\lambda(\Lambda \sim \Phi_R)}{8}\Phi_R^4 \tag{A2}$$

with the running effective parameters $m_H^2$ and $\lambda$. The connection with effective action at zero external momentum (i.e. the effective potential $V_{eff}$) is then

$$V_{eff}(\Phi_{cl}) = V(\Phi_R) \tag{A3}$$

where $\Phi_{cl}$ is the classical field (on which the generating functional of 1-Particle-Ireducible Green functions depends) corresponding to the running field $\Phi_R$. Obviously, it is the zero-temperature effective potential $V_{eff}$, and not some particular values of the running effective parameters $m_H^2$ and $\lambda$, that defines the vacuum structure of the theory. If the minimum of $V_{eff}$ is away from zero, the electroweak symmetry is broken.

The original Veltman's approach [56] can be used to describe the running effective potential in Eq. (A2). Veltman reached the conclusion that by redefining mass terms and fields (i.e. running), the SM Lagrangian with one-loop corrections (and with only quadratic divergences considered) may be brought in a gauge invariant fashion to the same form as the tree level Lagrangian. This result was confirmed by Osland and Wu [67] and refined with additional logarithmic terms at the one-loop level by Ma [68] in $R_\xi$. gauge. Moreover, the running of all the couplings of interest was included [150]. In addition, the higher-loop contributions to quadratic running were calculated in recursive manner [151, 152].

The running effective Higgs mass squared at one-loop level satisfies [66, 68]

$$\frac{dm_H^2}{d\Lambda^2} = \frac{3g_W^2}{64\pi^2 M_W^2}\left(m_H^2 + 2M_W^2 + M_Z^2 - 4\sum_f \frac{n_f}{3} m_f^2\right) \quad \text{(A4)}$$

$$+ \frac{3g_W^2}{64\pi^2 M_W^2} \frac{m_H^2}{2\Lambda^2}\left(m_H^2 - 2M_W^2 - M_Z^2 + 2\sum_f \frac{n_f}{3} m_f^2\right),$$

where $n_f = 3$ (1) for quarks (leptons). First term corresponds to the famous quadratic divergence. Using the tree level SM relations, the first term may be rewritten in terms of the gauge couplings $g_Y, g_W$, Yukawa couplings $g_f$'s, and quartic coupling $\lambda$ as

$$\frac{dm_H^2}{d\Lambda^2} = -\frac{1}{32\pi^2}\left(12g_t^2 - 6\lambda - \frac{9}{2}g_W^2 - \frac{3}{2}g_Y^2\right) + \cdots . \quad \text{(A5)}$$

In 2D theory, where both space-time dimensions and Dirac trace were set to 2 (they were both set to 4 in the original calculation [66]), the Higgs mass running in the leading order [33] is proportional to

$$\frac{dm_H^2}{d\Lambda^2} \propto \left(3m_H^2 + 2M_W^2 + M_Z^2 - 6\sum_f \frac{n_f}{3}m_f^2\right), \quad \text{(A6)}$$

$$\frac{dm_H^2}{d\Lambda^2} \propto \left(3g_t^2 - 3\lambda - \frac{3}{4}g_W^2 - \frac{1}{3}g_Y^2\right),$$

with logarithmic running in the leading order.

Conversely, in the $\overline{MS}$ scheme (see below), the scalar mass squared $m^2(t)$ in 4D is running just logarithmically! And the scalar mass squared should not be confused with running effective Higgs mass!

Does this mean that the SM in the $\overline{MS}$ scheme does not suffer from the quadratic divergences embodied in the running effective Higgs mass squared?

The answer is no, and to show this, it is useful to consider the effective potential $V_{eff}$. Following the approach in [56-58] in the $\overline{MS}$ scheme and in the 't Hooft-Landau gauge, the renormalization group improved one-loop effective potential [63-65] is

$$V_{eff} = V_0 + V_1 \text{ where } V_0 = -\frac{1}{2}m^2(t)\Phi_R^2(t) + \frac{1}{8}\lambda(t)\Phi_R^4(t) \quad \text{(A7)}$$

and $V_1 = \sum_{i=1}^5 \frac{n_i}{64\pi^2}[k_i\Phi_R^2(t) - k_i']^2\left[\log\frac{k_i\Phi_R^2(t)-k_i'}{\mu^2(t)} - c_i\right] + \boldsymbol{\Omega(t)}.$

The values of the parameters $n, k, k',$ and $c$ here are same as in [56-58]. Contribution to the cosmological constant is denoted by $\boldsymbol{\Omega}$ and is assumed irrelevant for the current calculation (though it may have huge importance for the physics of early Universe [70]).

Classical and running Higgs fields are related as

$$\Phi_R(t) = exp\left[-\int_0^t \gamma(t')dt'\right]\Phi_{cl} , \quad \text{(A8)}$$

where $\gamma(t') = \frac{3}{16\pi^2}\left[g_t^2(t') - \frac{1}{4}g_Y^2(t') - \frac{3}{4}g_W^2(t')\right]$

is the anomalous dimension, and $g_t$ is the top Yukawa coupling. The boundary condition for the $\overline{MS}$ scheme's mass squared, $m^2(t)$, is obtained by requiring $v_{EW} = 2.462 \cdot 10^2 \, GeV$; see Equ (14) in [56].

By using Equ (A2-A3) the *running effective* Higgs mass squared is extracted from Eqs. (A7-A8) as

$$m_H^2 = -\frac{4V_{eff}(\Phi_{cl}) - \frac{\lambda}{2}\Phi_R^4}{\Phi_R^2} \, . \tag{A9}$$

This is a quantity in the $\overline{MS}$ scheme that should be compared with the quadratically unstable Higgs mass squared, Eq. (A4), in the method developed from the original Veltman's approach!

As discussed in [26, 33] results are identical in both the SM dimensional $\overline{MS}$ regularization and in the Veltman's hard-cutoff method, the two most popular and most reliable approaches, to a very high precision with relatively small numerical processing error.

In this study the running Higgs masses squared as obtained from the $\overline{MS}$ effective potential approach [63-65] as well as from the Euclidean hard cut-off generalized Veltman's approach are analyzed in the similar manner: at the one-loop level with the logarithmic terms included and with running of all the couplings of interest at the two-loop level, i.e. in the next-to-leading-log (NTLL) level approximation.

In the original study [26, 33] the strong coupling and the top pole mass were $\alpha_S = 0.1182$ and $m_t = 175 \, GeV$ respectively. Here, the results are recalculated for the current world average value $m_t = 173.1 \, GeV$ [69].

The matching condition for the running top Yukawa coupling is identical to the one in [56-58]. The one-loop level matching conditions for $m^2$ and $\lambda$ in the $\overline{MS}$ effective potential approach, and $m_H^2$ and $\lambda$ in the hard cut-off generalized Veltman's approach were obtained from the standard requirement that

$$V_{eff}^{(1)}|_{\langle\Phi\rangle} = 0 \quad \text{and} \quad V_{eff}^{(2)}|_{\langle\Phi\rangle} = m_H^2(0) = m_H^2(pole) - \Delta\Pi\left(m_H^2(pole)\right) \tag{A10}$$

where $\Delta\Pi\left(m_H^2(pole)\right) = Re[\Pi\left(m_H^2(pole)\right) - \Pi(0)]$ with $\Pi$ being the renormalized self-energy of the Higgs boson. The reader is directed to reference [56,57] for more details. That approach has been closely followed here with the main results reconfirmed with a very good precision.

In the case of the Euclidean hard cut-off generalized Veltman's approach, the main difference in the matching procedure is in different form of the running effective potential Eq. (A2). The running of the Higgs mass squared is given by Eq. (A4-A5). Although lengthy, the matching procedure is rather trivial.

After the matching conditions are properly set, the gauge and Yukawa (top is only relevant) couplings are set to run at the two-loop level [63, 153, 154].

The variations in the physical Higgs mass upper bound as obtained in [26] from the existence of both HMZC scales are due to the variation in $\alpha_S$ and $m_t$ in the linear approximation as

$$\delta m_H[GeV] \cong 1.40 \, \delta m_t[GeV] - 360 \delta\alpha_S \quad \text{and} \tag{A11}$$

$$\delta m_H[GeV] \cong 1.80\, \delta m_t[GeV] - 70\delta\alpha_S$$

in the $\overline{MS}$ scheme and generalized Veltman's method, respectively. The errors are determined in a rather conservative manner. They are obtained from the separate variations of the matching conditions, in the range from the one-loop level matching to the tree level matching, for the three main parameters: $m_H^2$ (or $m^2$), $\lambda$ and $g_t$. Top quark Yukawa coupling matching condition is mainly responsible for the upper errors. The quartic coupling matching condition tends to bring the results of the two methods closer. The results are rather insensitive to the variations in the mass squared matching conditions. Following the same logic as in [56-58] for the $\overline{MS}$ scheme, the O(3 GeV) uncertainty is also incorporated in response to the requirement that the effective potential $V_{eff}$, must be renormalization scale independent. The separate errors on the physical Higgs mass have been added in quadrature.

Here I used two completely different regularization methods and I show that the results are essentially identical. How is that possible?

Once again, it is important to make a distinction between the $\overline{MS}$ parameters $m$ and $m_H$. The $\overline{MS}$ mass parameter $m$, has intrinsic logarithmic running. Whereas $m_H$ is a quantity derived from the $\overline{MS}$ parameter $m$ and obtained from the one loop corrections to the effective potential. And it is the parameter $m_H$ defined by the tree-level form of the effective potential which runs quadratically.

In the hard cutoff Euclidean regularization scheme the integrals of the type

$$\int d^d p, \quad \int \frac{d^d p}{p^2}, \quad \ldots \tag{A12}$$

are nonzero while in dimensional regularization due to the dilatation property they are identically zero. As can be shown the two regularization methods however agree in the logarithmic terms. This interplay may be easily seen if one dimensionally continues and regularizes propagators by the method of Pauli-Villars [155]. Then one finds [94] for example for d < 4

$$\int \frac{d^d q}{(2\pi)^2} \frac{1}{q^2(p+q)^2} \sim \frac{(p^2)^{(d/2)-2} - \Lambda^{d-4}}{8\pi^2(4-d)} \tag{A13}$$

Fixing the cutoff and taking the limit d=4 one obtains $ln(\Lambda)$ instead of a pole at d-4. Vice versa, by fixing d-4 and taking the cutoff to infinity one obtains the continuation of the initial integral.

## Acknowledgement

I am indebted to Gorazd Cvetic, Thomas Hambye, and Chris T. Hill, for valuable comments and encouragements to publish this manuscript. I thank Bo Morgan and Tijana Vujosevic for editing part of this manuscript. I am also indebted to many other great people who taught me how to think about the physics and beyond; several of them are listed below.

# References

[1] S. L. Glashow, Nucl. Phys. 22, 579-588. (1961)

[2] S. Weinberg, Phys. Rev. Lett. 19, 1264-1266. (1967)

[3] A. Salam, Proc. of the 8th Nobel Symposium on `Elementary Particle Theory, Relativistic Groups and Analyticity', Stockholm, Sweden, 1968, edited by N. Svartholm, p. 367-377. (1969)

[4] G. 't Hooft, Nucl. Phys. B33, 173-199. (1971)

[5] G. 't Hooft, Nucl. Phys. B35, 167-188. (1971)

[6] G. 't Hooft and M. J. G. Veltman, Nucl. Phys. B44, 189-213. (1972)

[7] D. J. Gross and F. Wilczek, Phys. Rev. Lett. 30, 1343-1346. (1973)

[8] D. H. Politzer, Phys. Rev. Lett. 30, 1346-1349. (1973)

[9] D. J. Gross and F. Wilczek, Phys. Rev. D8, 3633-3652. (1973)

[10] D. J. Gross and F. Wilczek, Phys. Rev. D9, 980-993. (1974)

[11] D. H. Politzer, Phys. Rep. 14 (1974) 129-180.

[12] C. N. Yang and R. L. Mills, Phys. Rev. 96, 191-195. (1954)

[13] J. Goldstone, Nuovo Cim. 19, 154-164. (1961)

[14] J. Goldstone, A. Salam, and S. Weinberg, Phys. Rev. 127, 965-970. (1962)

[15] P. W. Higgs, Phys. Lett. 12, 132-133. (1964)

[16] F. Englert and R. Brout, Phys. Rev. Lett. 13, 321-322. (1964)

[17] G. S. Guralnik, C. R. Hagen, and T. W. B. Kibble, Phys. Rev. Lett. 13, 585-587. (1964)

[18] J. D. Bjorken and S. L. Glashow, Phys. Lett. 11, 255-257. (1964)

[19] P. W. Higgs, Phys. Rev. 145, 1156-1163. (1966)

[20] T. W. B. Kibble, Phys. Rev. 155, 1554-1561. (1967)

[21] S. Weinberg, Phys. Rev. Lett. 27, 1688-1691. (1971)

[22] A. Einstein, Annalen der Physik 17: 891–921. (1905)

[23]CERN Press release

 <http://press.web.cern.ch/Press/PressReleases/Releases2000/PR08.00ELEPRundelay.html>  (2000)

[24] P. Renton, Nature 428, 141-144 (2004)

[25] LHC website <http://public.web.cern.ch/public/en/LHC/LHC-en.html>  (2010)

[26] M. B. Popovic, Boston University High Energy Physics, BUHEP-01-15, hep-ph/0106355 (2001)

[27] J. Wess and B. Zumino, Nucl. Phys. B70, 39. (1974)

[28]  H. P. Nilles, Phys. Reports, **110**, 1 (1984).

[29]  H. E. Haber and G. L. Kane, Phys. Reports, **117**, 75 (1985).

[30] J. Wess and J. Bagger, Supersymmetry and Supergravity, Princeton University Press. (1983)

[31] S. Ferrara ed., Supersymmetry , North Holland and World Scientific. (1987)

[32] C. T. Hill, FERMILAB-PUB-80-097-T, Dec 1980. 44pp. Published in Phys.Rev.D24, 691. (1981)

[33] M. B. Popovic, Harvard University Theory Physics, HUTP-02/A012, hep-ph/0204345 (2002)

[34]  S. Coleman and E. Weinberg, Phys. Rev. D7, 1888. (1973)

[35] B. W. Lee, C. Quigg, and H.B. Thacker, Phys. Rev. D16, 1519. (1977)

[36]  C. Quigg, in "Flavor Physics for the Millennium TASI 2000", ed. J. L. Rosner, World Scientific. (2000)

[37]  A. Einstein, "Die Feldgleichungen der Gravitation", Sitzungsberichte der Preussischen Akademie der Wissenschaften zu Berlin: 844–847. (1915)

[38]  A. Einstein, Annalen der Physik 49, (1916)

[39]  K. G. Wilson and J. Kogut, Phys. Reports, 12C, 75. (1974)

[40]  L. Susskind, Phys. Rev. D20, 2619. (1979)

[41]  F. Wilczek, "What is space?" talk at MIT, March 2010. (2010)

[42]  J. Hogan, Nature 448 (7151), 240–245. (2007)

[43] M. P. Hobson, G. P. Efstathiou, and A. N. Lasenby (2006), "General Relativity: An introduction for physicists" (Reprinted with corrections 2007 ed.). Cambridge University Press. p. 187. (2007)

[44]  ALEPH Collab., Phys. Lett. B526, 191. (2002)

[45]  ALEPH, DELPHI, L3, and OPAL Collaborations, The LEP Working Group for Higgs Boson Searches, Phys. Lett. B565, 61. (2003)

[46]  C. Amsler et al. (Particle Data Group), Physics Letters B667, 1 (2008) and 2009 partial update for the 2010 edition (URL: http://pdg.lbl.gov). (2010)

[47]   LEP Electroweak Working Group, status of September 2007, <http://lepewwg.web.cern.ch/LEPEWWG/>;  M. Grunewald, arXiv:0709.3744v1(2007);  J. Erler and P. Langacker, Electroweak Model and Constraints on New Physics  in Particle Data Group (2008)

[48]   H. Flacher, M. Goebel, J. Haller, A. Hocker, K. Moenig and J. Stelzer, Eur. Phys. J. C60, 543. (2009)

[49]   T. Aaltonen et al., (CDF and D0 Collaborations), Phys. Rev. Lett. 104, 061802. (2010)

[50]   T. Aaltonen et al., (CDF Collaboration), Phys. Rev. Lett. 104, 061803. (2010)

[51]   V. M. Abazov et al. (D0 Collaboration), Phys. Rev. Lett. 104, 061804. (2010)

[52]   N. Cabibbo, L. Maiani, G. Parisi, and R. Petronzio, Nucl. Phys. B158, 295. (1979)

[53]   M. Lindner, Z. Phys. C31, 295. (1986)

[54]   M. Sher, Phys. Rept. 179, 273. (1989)

[55]   G. Altarelli and G. Isidori, Phys. Lett. B337, 141. (1994)

[56]   J. A. Casas, J. R. Espinosa, and M. Quiros, Phys. Lett. B342, 171. (1995)

[57]   J. A. Casas, J. R. Espinosa, and M. Quiros, Phys. Lett. B382, 374. (1996)

[58]   M. Quiros, in "Perspective on Higgs Physics II", G. Kane, ed., World Scientific, Singapore. (1998)

[59]   Yu. F. Pirogov and O. V. Zenin, Eur. Phys. J. C10, 629. (1999)

[60]   P. Q. Hung and G. Isidori, Phys. Lett. B402, 122. (1997)

[61]   T. Hambye and K. Riesselmann, Phys. Rev. D55, 7255. (1997)

[62]   G. 't Hooft and M. Veltman, Nucl. Phys. B61, 455 (1973); W. A. Bardeen, A. J, Buras, D. W. Duke and T. Muta, Phys. Rev. D18, 3998. (1978)

[63]   C. Ford, D. R. T. Jones, P. W. Stephenson and M. B. Einhorn, Nucl. Phys. B395, 17. (1993)

[64]   M. Bando, T. Kugo, N. Maekawa and H. Nakano, Phys. Lett. B301, 83. (1993)

[65]   M. Bando, T. Kugo, N. Maekawa and H. Nakano, Prog. Theor. Phys. 90, 405. (1993)

[66]   M. Veltman, Acta Phys. Pol. B12, 437. (1981)

[67]   P. Osland and T. T. Wu, Z. Phys. C55, 585. (1992)

[68]   E. Ma, Phys. Rev. D47, 2143. (1993)

[69]   arXiv:0903.2503v1 [hep-ex] (2009)

[70]   E. W. Kolb and M. S. Turner, "The Early Universe", Addison-Wesley Publishing Company. (1990)

[71]  R. Contino, Lecture at TASI 2009, arXiv:1005.4269v1 [hep-ph] (2010)

[72]  J. F. Gunion, H. E. Haber, G. Kane, and S. Dawson, "The Higgs Hunter's Guide," Addison-Wesley Publishing Company, Redwood City, CA. (1990)

[73]  S. Dawson, Four lectures given at NATO Advanced Study Institute on Techniques and Concepts of High Energy Physics, July 11-22, 1996, St. Croix, Virgin Islands [hep-ph/9612229]. (1996)

[74]  arXiv:hep-ex/0608032v1 [hep-ex] (2006)

[75]  M. Carena, J. R. Espinosa, M. Quiros, and C. E. M. Wagner, Phys. Lett. B335, 209. (1995)

[76]  M. Carena, M. Quiros, and C. E. M. Wagner, Nucl. Phys. B461, 407. (1996)

[77]  H. E. Haber, R. Hempfling, and A. H. Hoang, Z. Phys. C75, 539. (1997)

[78]  R. Barbieri and G. Giudice, Nucl. Phys. B306, 63. (1988)

[79]  American Linear Collider Working Group, Linear Collider Physics Resource Book for Snowmass 2001, section 2.4.1, pp 17-20. (2001)

[80]  S. Coleman, "Fate of the false vacuum: Semiclassical theory". Phys. Rev. D15: 2929–36. (1977)

[81]  C. Callan and S. Coleman, "Fate of the false vacuum. II. First quantum corrections". Phys. Rev. D16: 1762–68. (1977)

[82]  M. Stone, "Lifetime and decay of excited vacuum states". Phys. Rev. D 14: 3568–3573. (1976)

[83]  P.H. Frampton, "Vacuum Instability and Higgs Scalar Mass". Phys. Rev. Lett. 37: 1378–1380. (1976)

[84]  M. Stone, "Semiclassical methods for unstable states". Phys.Lett. B 67: 186–183. (1977)

[85]  P.H. Frampton, "Consequences of Vacuum Instability in Quantum Field Theory". Phys. Rev. D15: 2922–28. (1977)

[86]  S. Coleman and F. De Luccia, "Gravitational effects on and of vacuum decay". Physical Review D21: 3305. (1980)

[87]  A. H. Guth, "The Inflationary Universe: A Possible Solution to the Horizon and Flatness Problems". Phys. Rev. D23: 347. (1981)

[88]  S. W. Hawking and I. G. Moss, "Supercooled phase transitions in the very early universe". Phys. Lett. B110: 35–8. (1982)

[89] A. Linde, "A New Inflationary Universe Scenario: A Possible Solution Of The Horizon, Flatness, Homogeneity, Isotropy And Primordial Monopole Problems". Phys. Lett. B108: 389. (1982)

[90] M.S. Turner and F. Wilczek, "Is our vacuum metastable?" Nature 298: 633–634. (1982)

[91] P. Hut and M.J. Rees, "How stable is our vacuum?" Nature 302: 508–509. (1983)

[92] LHC Safety Assessment Group Report, <http://cern.ch/lsag/LSAG-Report.pdf>

[93] J. Schwinger, Phys. Rev. 128, 2435. (1962)

[94] J. Zinn-Justin, "Quantum Field Theory and Critical Phenomena" 2nd ed., Clarendon Press, Oxford (1993).

[95] C. M. Sommerfield, Annals Phys. 26 1-43. (1963)

[96] W. E. Thirring and J. E. Wess, Annals Phys. 27 331-337. (1964)

[97] H. Georgi, Phys. Rev Lett. 98, 221601. (2007)

[98] H. Georgi and Y. Kats, Phys. Rev. Lett. 101, 131603. (2008)

[99] C. T. Hill and E. H. Simmons, Physics Reports 381: 235–402. (2003)

[100] G. Cvetic, "Top quark condensation," Rev. Mod. Phys. 71 513-574. (1999)

[101] J. Bardeen, L. N. Cooper, and J. R. Schrieffer, Phys. Rev. 106, 162 – 164. (1957)

[102] Y. Nambu and G. Jona-Lasinio, Physical Review 122: 345–358. (1961)

[103] Y. Nambu and G. Jona-Lasinio, Physical Review 124: 246–254. (1961)

[104] Y. Nambu, in Z. Adjduk, S. Pokorski, and A. Trautman. Proceedings of the Kazimierz 1988 Conference on New Theories in Physics. XI International Symposium on Elementary Particle Physics. pp. 406–415. (1989)

[105] Vaks V. G., Larkin A. I., "On the application of the methods of superconductivity theory to the problem of the masses of elementary particles," Zh. Eksp. Theor. Fiz. 40 282-285. (1961) [English transl. : Sov. Phys. JETP 13 192-193. (1961)]

[106] V.A. Miransky, M. Tanabashi, and K. Yamawaki, DPNU-89-08, Jan 1989. 20pp. Published in Mod.Phys.Lett.A4:1043. (1989)

[107] W. A. Bardeen, C. T. Hill, and M. Lindner, FERMILAB-PUB-89-127-T, Jul 1989. 43pp. Published in Phys.Rev.D41, 1647. (1990)

[108] S. Weinberg, Phys. Rev. D 13, 974. (1976)

[109] S. Weinberg, Phys. Rev. D 19, 1277. (1979)

[110] A. Hasenfratz, P. Hasenfratz, K. Jansen, J. Kuti and Y. Shen, Nucl. Phys. B 365, Issue 1, 79-97. (1991)

[111] Y. Nambu, "Bootstrap Symmetry Breaking in Electroweak Unification," Enrico Fermi Institute Preprint, 89-08. (1989).

[112] G. Cvetic, "Top quark effects on the scalar sector of the minimal standard model," Int. J. Mod. Phys. A11 5405. (1996)

[113] T. Hambye, "Symmetry breaking induced by top quark loops from a model without scalar mass," Phys. Lett. B371, 87. (1996)

[114] H. Pagels and S. Stokar, Phys. Rev. D 20, 2947. (1979)

[115] W. A. Bardeen, C. N. Leung, and S. T. Love, Phys. Rev. Lett. 56, 1230. (1986)

[116] C. N. Leung, S. T. Love and W. A. Bardeen, Nucl. Phys. B273, 649. (1986)

[117] C. T. Hill, Phys.Lett.B266, 419-424. (1991)

[118] C. T. Hill Phys. Lett. B345, 483-489, (1995)

[119] K. Lane and E. Eichten, Phys. Lett. B 352, 382. (1995)

[120] K. D. Lane, Phys. Rev. D 54, 2204. (1996)

[121] G. Buchalla, G. Burdman, C. T. Hill, and D. Kominis, Phys. Rev. D 53, 5185. (1996)

[122] R. S. Chivukula, E. H. Simmons, and J. Terning, Phys. Lett. B331, 383. (1994)

[123] R. S. Chivukula, E. H. Simmons, and J. Terning, Phys. Rev. D53, 5258. (1996)

[124] E. H. Simmons, Phys. Rev. D55, 5494. (1997)

[125] R. S. Chivukula, A. G. Cohen and E. H. Simmons, Phys. Lett. B380, 92. (1996)

[126] M. B. Popovic and E. H. Simmons, Phys. Rev. D 58, 095007. (1998)

[127] B. A. Dobrescu and C. T. Hill, Phys. Rev. Lett. 81, 2634. (1998)

[128] R. S. Chivukula, B. A. Dobrescu, H. Georgi, and C. T. Hill, Phys. Rev. D 59, 075003. (1999)

[129] M.B. Popovic, Phys. Rev. D 64, 035001. (2001)

[130] N. Arkani-Hamed, H.-C. Cheng, B. A. Dobrescu, and L. J. Hall, Phys. Rev. D 62, 096006. (2000)

[131] T. Appelquist, H.-C. Cheng, and B. A. Dobrescu, Phys. Rev. D 64, 035002. (2001)

[132] S. Dimopoulos and L. Susskind, Nucl. Phys. B155, 237. (1979)

[133] E. Eichten and K. Lane, Phys. Lett. 90B, 125. (1980)

[134] R. S. Chivukula, S. B. Selipsky, and E. H. Simmons, Phys. Rev. Lett. 69, 575. (1992)

[135] T. Appelquist, M. J. Bowick, E. Cohler, and A. I. Hauser, Phys. Rev. Lett. 53, 1523. (1984)

[136] M. E. Peskin and D. V. Schroeder, An Introduction to Quantum Field Theory, Addison Wesley Publishing Company. (1995)

[137] S. Bethke, arXiv:0908.1135 [hep-ph]. (2009)

[138] E.P. Wigner, Annals of Mathematics, 40, 149. (1939)

[139] L. H. Ryder, Quantum Field Theory, 2nd edition, Cambridge University Press. (1996)

[140] G. 't Hooft, Naturalness, chiral symmetry, and spontaneous chiral symmetry breaking. Lecture given at Cargese Summer Inst., Cargese, France, Aug 26 - Sep 8, 1979. (1979)

[141] N. Arkani-Hamed, A. G. Cohen, E. Katz, and A. E. Nelson, Harvard University Theory Physics, HUTP-02/A017, hep-ph/0206021, JHEP 07, 034. (2002)

[142] N. Arkani-Hamed, A. G. Cohen, and H. Georgi, Phys. Lett. B513, 232–240. (2001)

[143] N. Arkani-Hamed, A.G. Cohen, T. Gregoire, E. Katz, A.E. Nelson, and J.G. Wacker, JHEP 08, 021. (2002)

[144] M. Schmaltz, Nucl. Phys. Proc. Suppl. 117 40–49. (2003)

[145] D. B. Kaplan, H. Georgi, and S. Dimopoulos, Phys. Lett. B136 187. (1984)

[146] H. Georgi, D. B. Kaplan, and P. Galison, Phys. Lett. B143 152. (1984)

[147] H. Georgi and D. B. Kaplan, Phys. Lett. B145 216. (1984)

[148] M. J. Dugan, H. Georgi, and D. B. Kaplan, Nucl. Phys. B254 299. (1985)

[149] K. Lane and A. Martin, Phys. Lett. B635, 118-122. (2006)

[150] M. Chaichian, R. G. Felipe and K. Huitu, Phys. Lett. B363, 101. (1995)

[151] M. Einhorn and D. R. T. Jones, Phys. Rev. D46, 5206. (1992)

[152] C. Kolda and H. Murayama, JHEP 0007, 035. (2000)

[153] M. E. Machacek and M. T. Vaughn, Nucl. Phys. B222, 83. (1983)

[154] M. E. Machacek and M. T. Vaughn, Nucl. Phys. B236, 221 (1984)

[155] W. Pauli and F. Villars, Rev. Mod. Phys. 21, 434. (1949)